\documentclass{aastex61}

\usepackage{amsmath}
\usepackage{graphicx}
\usepackage{multirow}
\usepackage{color}
\usepackage{natbib}
\usepackage{morefloats}

\begin{document}

\title{The curiously warped mean plane of the Kuiper belt}

\author[0000-0001-8736-236X]{Kathryn Volk}
\correspondingauthor{Kathryn Volk}
\email{kvolk@lpl.arizona.edu}
\affil{{Lunar and Planetary Laboratory, The University of Arizona, 1629 E University Blvd, Tucson, AZ 85721}}
\author[0000-0002-1226-3305]{Renu Malhotra}
\affil{{Lunar and Planetary Laboratory, The University of Arizona, 1629 E University Blvd, Tucson, AZ 85721}}

\begin{abstract}
We measured the mean plane of the Kuiper belt as a function of semi-major axis. 
For the classical Kuiper belt as a whole (the non-resonant objects in the semi-major axis range 42--48~au), we find a mean plane of inclination $i_m=1.8^{\circ}$$^{+0.7^{\circ}}_{-0.4^{\circ}}$ and longitude of ascending node $\Omega_m=77^{\circ}$$^{+18^{\circ}}_{-14^{\circ}}$ (in the J2000 ecliptic-equinox coordinate system), in accord with theoretical expectations of the secular effects of the known planets. 
With finer semi-major axis bins, we detect a statistically significant warp in the mean plane near semi-major axes 40--42~au.  Linear secular theory predicts a warp near this location due to the $\nu_{18}$ nodal secular resonance, however the measured mean plane for the 40.3-42~au semi-major axis bin (just outside the $\nu_{18}$) is inclined $\sim13^{\circ}$ to the predicted plane, a nearly 3-$\sigma$ discrepancy.
For the more distant Kuiper belt objects of semi-major axes in the range 50--80~au, the expected mean plane is close to the invariable plane of the solar system, but the measured mean plane deviates greatly from this: it has inclination $i_m=9.1^{\circ}$$^{+6.6^{\circ}}_{-3.8^{\circ}}$ and longitude of ascending node $\Omega_m=227^{\circ}$$^{+18^{\circ}}_{-44^{\circ}}$.  We estimate this deviation from the expected mean plane to be statistically significant at the $\sim97-99\%$ confidence level.  We discuss several possible explanations for this deviation, including the possibility that a relatively close-in  ($a\lesssim100$~au), unseen small planetary-mass object in the outer solar system is responsible for the warping. 
\end{abstract}

\section{Introduction}\label{s:intro}
Over the past two decades, discoveries of minor planets in the outer solar system have revealed complex dynamical features and prompted new theoretical models of the formation and evolution of the solar system. 
One of the most surprising findings is that the orbital planes of Kuiper belt objects (KBOs) are widely dispersed.  
While many investigators have remarked on the wide dispersion of KBOs' orbital inclinations, only a few have attempted to accurately measure their mean plane.  
It might be supposed {\it a priori} that the mean plane of the Kuiper belt should be close to the mean plane of the solar system itself.  
This plane, also known as the ``invariable plane", is normal to the total orbital angular momentum of the solar system; it has been determined to better than one milliarcsecond accuracy based on the eight planets, Mercury--Neptune, plus the dwarf planets Ceres and Pluto and the two largest asteroids, Vesta and Pallas \citep{Souami:2012}.  
Unseen large masses in the outer solar system would affect the mean plane of the solar system, especially if those masses are on significantly inclined orbits \citep[e.g.][]{Goldreich:1972}. 
Therefore, accurately measuring the mean plane of the observed Kuiper belt objects offers a potentially sensitive probe of unseen planetary mass objects beyond Neptune.

The earliest attempt to determine the mean plane of the observed Kuiper belt appears to have been by \cite{Collander-Brown:2003}. 
These authors defined several subsets of the known KBOs that might best define the mean plane, focusing on the so-called classical KBOs with semi-major axes in the range 40--47~au; they concluded that the average orbital angular momentum of the classical KBOs (a sample of 141 at the time) was consistent with their mean plane being very close to the invariable plane, but they did not assess the error in their measurement.  
Measuring the mean plane of an ensemble of minor planets' orbits by averaging their angular momentum vectors (in practice, averaging the unit orbit normal vectors) is susceptible to serious systematic errors due to observational biases. 
\cite{Brown:2004b} used a more robust method of finding the mean plane by determining the plane of symmetry of the KBOs' sky-plane motion vectors. 
They applied this method to all 728 KBOs then known (of median semi-major axis $a=44$~au) and concluded that the Kuiper belt's mean plane is not consistent with the invariable plane but is instead consistent with the local Laplacian plane (i.e., the plane forced by the secular effects of the planets) at $a=44$~au predicted by linear secular perturbation theory.  
A similar detailed study by \cite{Elliot:2005} reached a different conclusion, finding that the Kuiper belt's mean plane is more consistent with the invariable plane than with the local Laplacian plane at $a=44$~au. 
A subsequent study by \cite{Chiang:2008} pointed out that the local Laplacian plane of the classical Kuiper belt should be warped by several degrees near $a=40.5\pm1$~au; this warp is owed to the $\nu_{18}$ nodal secular resonance, which is driven mainly by Neptune. 

Since the time of the previous studies, the number of known Kuiper belt objects has more than doubled, and the new discoveries now encompass a larger range of semi-major axes, motivating a re-examination of the Kuiper belt's plane.  
We are also motivated by recent intriguing but not conclusive evidence of a large unseen planet on an inclined orbit in the distant solar system~\citep{Trujillo:2014,Batygin:2016,Brown:2016,Malhotra:2016,Holman:2016a,Holman:2016b,Sheppard:2016} as well as previous suggestions of unseen planetary mass objects perturbing the orbits of Kuiper belt objects \citep[e.g.,][]{Gladman:2002,Lykawka:2008}. Unseen planetary mass objects in the outer solar system could result in KBOs having a mean plane that deviates from that expected due to the known planets.

We are additionally motivated by the desire to have a more complete and accurate model for the overall distribution of KBO orbital planes. 
This distribution has important implications for the dynamical history of the outer solar system. 
The observationally derived widths of the KBOs' inclination distribution function have been used as constraints on theoretical models of the dynamical excitation history of the outer solar system~\citep[e.g.,][]{Gomes:2003,Nesvorny:2015}. 
In the literature, the measures of the effective width of the inclination distribution function are based on KBOs' inclinations relative to the ecliptic plane \citep[e.g.,][]{Brown:2001}, relative to the invariable plane \citep[e.g.,][]{Petit:2011}, or relative to the mean plane of the entire Kuiper belt \citep[usually assumed to be consistent with the invariable plane, e.g.,][]{Elliot:2005,Gulbis:2010}. 
However, the most dynamically meaningful measure of the effective width of the inclination distribution will be that measured for inclinations relative to the local forced plane.  
If there are populations within the Kuiper belt where the local forced plane, i.e., the true mean plane, is significantly inclined (as we show in Section~\ref{s:results}), then the distribution of inclinations relative to the ecliptic or invariable planes could be a misleading measure of the true distribution of KBO orbital planes; it could lead to potentially incorrect conclusions when using the effective width of the inclination distribution as an observational test of Kuiper belt formation models. 
An accurate measurement of the Kuiper belt's mean plane as a function of semi-major axis will allow for a more dynamically meaningful representation of the distribution of KBO orbital planes.

In the present work, we analyze the current data to measure the Kuiper belt plane as a function of semi-major axis, and we compare the results with theoretical expectations.  
The paper is organized as follows. 
We briefly describe in Section~\ref{s:theory} the expected mean plane for the classical Kuiper belt based on standard linear secular theory and the expected mean plane for the scattered disk based on symmetry arguments.  
The dataset of KBO orbits is described in Section~\ref{s:data}, with greater detail given in Appendix~\ref{a:data}. We describe in Section~\ref{s:methods} the methods we use to measure the mean plane of KBOs and how we estimate the associated uncertainties.   
The results of these calculations are presented in Section~\ref{s:results}. 
In Section~\ref{s:disc}, we summarize our findings and discuss their implications.

\section{Expected mean plane}\label{s:theory}

The orbital distribution of minor planets in the solar system is strongly influenced by long-term perturbations from the planets. 
The planetary effects can be complicated in many ways, but the mean plane enforced by their long-term perturbations can be identified with the so-called ``forced inclination vector", $(q_0,p_0) = (\sin i_0\sin\Omega_0,\sin i_0\cos\Omega_0)$, which is the homogeneous solution to the linear secular perturbation equations for a minor planet. 
This is relatively simply calculated in the Laplace-Lagrange linear secular theory~\citep{Murray:1999SSD},
\begin{eqnarray}
(q_0,p_0) &=& (\sin i_0\cos\Omega_0,\sin i_0\sin\Omega_0) \\ \nonumber
&=& \sum_{i=1}^{8} \frac{\mu_i}{f_i-f_0} (\cos\gamma_i,\sin\gamma_i),
\end{eqnarray} 
where $f_0$ is the nodal precession rate induced by the orbit-averaged quadrupolar potential of the planets, $f_i$ and $\gamma_i$ are the frequencies and phases of the secular modes of the solar system's eight major planets, and $\mu_i$ is the weighting factor for each secular mode;  $f_i$ and $\gamma_i$ depend only upon the planetary parameters, while $f_0$ and $\mu_i$ depend additionally on the minor planet's semi-major axis. (Following standard practice, we use the heliocentric coordinate system of the J2000.0 ecliptic-equinox throughout this paper, unless indicated otherwise.)
We note that the forced inclination vector, $(q_0,p_0)$, is related to the ecliptic projection of the unit vector, $\hat{\bf n}_0=(\sin i_0\sin\Omega_0,-\sin i_0\cos\Omega_0,\cos i_0)$, which is normal to the local Laplacian plane.
Over secular timescales, as the planets' orbits evolve under their own mutual perturbations, the forced inclination vector changes. Even so, for a population of minor planets with some dispersion in semi-major axes and orbital planes, the mean plane will coincide with the forced plane at that epoch~\citep{Chiang:2008}.

The forced inclination and longitude of ascending node, $i_0$ and $\Omega_0$, respectively, of a test particle are a function of its semi-major axis.  
We plot $i_0$ and $\Omega_0$ for the semi-major axis range 30--150~au in Figure~\ref{f:ckbi}, according to the linear secular solution in \cite{Murray:1999SSD}. 
This defines the local Laplacian plane as a function of semi-major axis, and is also the theoretically expected mean plane of non-resonant KBOs, as determined by the known major planets, Mercury--Neptune.
We observe that for large semi-major axes, $a\gg40$~au, the expected mean plane asymptotically approaches the invariable plane; the latter has inclination and longitude of ascending node of $1.58^{\circ}$ and $107.6^{\circ}$, respectively~\citep{Souami:2012}. 
We also observe that the expected mean plane in the Kuiper belt is not flat: it has a prominent warp (of several degrees) near $a=40.5$~au, owed to the $\nu_{18}$ nodal secular resonance. 
\cite{Chiang:2008} attempted to measure this warp in the observed KBOs with partial success. 
In the present work, we re-visit this problem with the larger observational sample now available.  
We also examine whether the mean plane of the more distant KBOs is consistent with the invariable plane.

The secular solution for the forced inclination vector of a minor planet is formally valid in the approximation of small inclinations. 
The minor planet's semi-major axis must also be well separated from both the planets' semi-major axes and from strong mean motion resonances with the planets. 
Because secular perturbation theory considers only orbit-averaged perturbations on the inclination vector, the minor planets' orbits must also not be planet-crossing.
These conditions are met fairly well in the classical Kuiper belt which consists of the non-resonant KBOs with $42\lesssim a/{\rm au}\lesssim48$. 
However, it is questionable whether the linear secular theory is applicable to the more distant KBOs.
The current observational sample of distant KBOs, with $a\gtrsim50$~au, is dominated by the ``scattering disk" and the ``scattered disk" populations.  
Many of these objects have current perihelion distances within a few au of Neptune's orbital radius, and are expected to gravitationally scatter with Neptune on $\sim10$ Myr timescales. This timescale is not much different from the secular precession timescale of their orbital planes (which ranges from a few Myr for $a\sim50$~au to several tens of Myr for $a\sim100$~au).  
In this circumstance, the Laplace-Lagrange secular theory is not a good description of the dynamics of these objects, so we must ask: what is the theoretically expected mean plane of the scattering and scattered disk objects?  
Provided that the gravitational scatterings with Neptune are not correlated over long timescales or amongst the KBOs, the plane of symmetry for the population of scattering and scattered objects must be close to the time-averaged orbital plane of Neptune, i.e., the invariable plane.  
There is no other preferred plane for this population, barring the effects of unseen distant masses.  
We examined the numerical simulations of the scattering population of KBOs that were carried out by \cite{Volk:2008} and \cite{Volk:2013}, and found that the mean plane of the simulated scattering KBOs remains close to the invariable plane (within $\sim1^{\circ}$ for 20 different subsets of $\sim250$ simulated scattering KBOs). 
This supports our argument above that the expected mean plane of the scattering and scattered population is the invariable plane.

\section{Kuiper belt observational data}\label{s:data}

\begin{figure*}[htbp]
   \centering
   \includegraphics[width=4.2in]{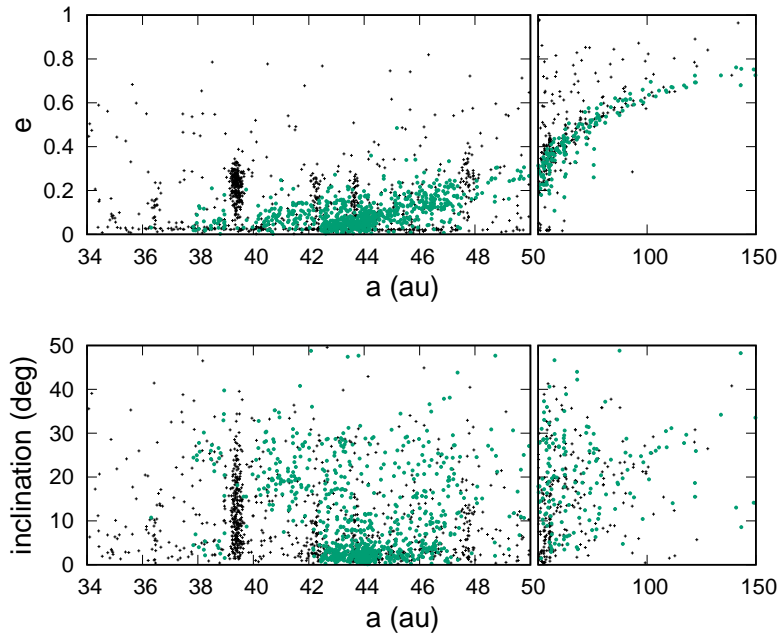}
   \caption{Eccentricity (top panel) and inclination (lower panel) vs. semi-major axis for the distant solar system objects listed in the Minor Planet Center (black crosses). The objects with $q>30$ au and $da/a<0.05$ that appear not to be in resonance with Neptune (our criteria for inclusion in the mean plane calculations) are shown as green dots. See Appendix~\ref{a:data} for additional information about these objects.}
   \label{f:a-e-i}
\end{figure*}

Our starting point is the list of minor planets in the outer solar system cataloged in the database of the Minor Planet Center\footnote{\url{http://www.minorplanetcenter.net/iau/lists/t\_centaurs.html} and \url{http://www.minorplanetcenter.net/iau/lists/t\_tnos.html}}, as of October 20, 2016. 
We gathered all the available astrometric observations for these objects and computed the best-fit barycentric orbit for each object using the \citet{Bernstein:2000} orbit fitting code. 
Figure~\ref{f:a-e-i} shows the semi-major axes, eccentricities, and inclinations for these objects (black crosses). 
We then selected those objects whose perihelion distances are beyond Neptune and whose semi-major axis uncertainty, $\delta a/a$, did not exceed 5\%.  
We numerically integrated their best-fit orbits forward for $10^7$ years under the influence of the Sun and the four giant planets to check for orbital resonances with Neptune. (We do this by checking for libration of a resonant angle for all mean motion resonances up to 30$^{th}$ order). 
We excluded objects in orbital resonances from further analysis because their orbital inclinations and precession rates are affected by resonant perturbations that are not described by secular theory. 
The remaining sample of objects is used in our Kuiper belt mean plane calculations; these are shown as green dots in Figure~\ref{f:a-e-i}. 
The complete listing of this sample, including their best fit orbital parameters and sky locations, is provided in Table~\ref{t:objects} in Appendix~\ref{a:data}.

\clearpage
\section{Methods}\label{s:methods}

\subsection{Measuring the mean plane}\label{s:mp}

A number of different methods have been used to measure the mean plane of the Kuiper belt.  
\cite{Collander-Brown:2003} used what might be considered the most intuitive method, namely computing the average of the unit vectors normal to the orbital planes of the observed sample of KBOs. 
The unit vector, $\hat{\bf h}$, normal to a KBO's orbit plane is expressed as
\begin{equation}
\hat{\bf h} = (\sin i\sin\Omega, -\sin i\cos\Omega,\cos i).
\label{e:hhat}\end{equation}
The inclination $i$ and longitude of ascending node $\Omega$ of a KBO's orbit can be accurately determined for even very short observational arcs.
However, measuring the mean plane as the average of $\hat{\bf h}$ is susceptible to serious systematic error when applied to an observationally biased sample.  
Consider, for example, a population of KBOs with a true mean plane inclined to the ecliptic and having an intrinsically large dispersion of inclinations to its mean plane. If the observed subset of this population were discovered only near the ecliptic (because that is where observational surveys were performed), then the observed sample would be systematically biased toward objects with orbit planes close to the ecliptic.  
The average of the unit orbit normal vectors for such a subset of objects would identify a mean plane with an inclination lower than the true mean plane.  
This and other systematic errors that can arise from averaging the unit orbit normal vectors are described in more detail in Appendix~\ref{a:biases}.

\cite{Brown:2004b} defined the mean plane as the plane of symmetry of the sky-plane motion vectors of KBOs, noting that on average the KBOs' sky-plane velocity vectors should be parallel to their plane of symmetry no matter where in the sky the KBOs are discovered.  
The unit vector in the direction of the sky-plane velocity of a KBO, $\hat{\bf v}_t$, can be determined from knowledge of its unit orbit normal and its sky-plane position.
The sky-plane position of a KBO is given by its ecliptic latitude and longitude, $\beta$ and $\lambda$. 
The unit vector directed along the heliocentric position of the KBO is then given by
\begin{equation}
\hat{\bf r} = (\cos\beta\cos\lambda, \cos\beta\sin\lambda,\sin\beta), 
\end{equation}  
and the sky-plane velocity vector has a direction given by $\hat{\bf v}_t= \hat{\bf h}\times\hat{\bf r}$.  
The ecliptic latitude and longitude of $\hat{\bf v}_t$ are given by
\begin{eqnarray}
\beta_{v_t} &=& \arcsin (\hat{\bf v}_t\cdot\hat{\bf z} ) \nonumber \\
&=& \arcsin[\sin i\cos\beta\cos(\lambda-\Omega)], \nonumber \\
\lambda_{v_t} &=& \arctan (\hat{\bf v}_t\cdot\hat{\bf y} /\hat{\bf v}_t\cdot\hat{\bf x}) \nonumber \\
 &=& \arctan \frac{\sin\beta\sin i\sin\Omega-\cos\beta\cos\lambda\cos i} {\sin\beta\sin i\cos\Omega + \cos\beta\sin\lambda\cos i}.
\end{eqnarray} 
We denote with $i_m$ and $\Omega_m$ the inclination and longitude of ascending node of the mean plane of the KBOs (where the subscript $m$ is to differentiate between the observationally derived mean plane and the theoretically expected mean plane, which has the subscript $0$).  The ecliptic latitude, $\beta_m$, at which this plane intersects the sky as a function of ecliptic longitude, $\lambda$, is given by
\begin{equation}
\beta_m(\lambda; i_m,\Omega_m) = \arcsin[\sin i_m \sin(\lambda-\Omega_m)].
\label{e:beta0}\end{equation}
\cite{Brown:2004b} measured the mean plane by minimizing the sum of the absolute value of the deviations of the KBOs' $(\beta_{v_t},\lambda_{v_t})$ from the curve defined by $\beta_m(\lambda)$ (Eq.~\ref{e:beta0}).
\cite{Elliot:2005} describe several variants of this method, including least squares minimization and maximum likelihood fits based on modeled distribution functions of the ecliptic latitudes $\beta_{v_t}$.  
In their implementation, they used the low inclination approximation of Eq.~\ref{e:beta0},
\begin{equation}
\beta_m(\lambda; i_m,\Omega_m) \approx  i _m\sin(\lambda-\Omega_m).
\label{e:beta00}\end{equation}

In general, computing the mean plane as the plane of symmetry of the sky-plane motion vectors should be more robust against systematic errors \citep[see section 2.1 in][]{Brown:2004b}.
We show in Appendix~\ref{a:biases} that this is indeed a much more robust method than averaging the orbit normal unit vectors, but it is not entirely free of systematic errors.  

Here we implement this approach in a computationally simpler and more efficient way than in previous studies.
Let $\hat{\bf n}$ denote the unit vector which is normal to the mean plane of the Kuiper belt.  
Then a KBO whose orbit plane coincides with the mean plane would have $\hat{\bf n}\cdot\hat{\bf v}_t=0$.  
A non-vanishing value of $\hat{\bf n}\cdot\hat{\bf v}_t$ would measure the deviation of the KBO's orbit plane from the mean plane.  
Therefore, to identify the mean plane of an observational sample of KBOs, we minimize the sum of the absolute values of $\hat{\bf n}\cdot\hat{\bf v}_t$ of all KBOs, $\sum|\hat{\bf n}\cdot\hat{\bf v}_t|$, over a grid of all possible unit vectors, $\hat{\bf n}$.

As a check, we confirmed that our method yields the same mean plane as \cite{Brown:2004b}'s method of minimizing $\sum |\beta_{vt}(\lambda_{vt})-\beta_m(\lambda_{vt})|$ when applied to the observational data of the classical KBOs (non-resonant objects in the semi-major axis range 42--48~au).

\subsection{Uncertainty of the measured mean plane}\label{s:uncertainties}

The measurement error of the computed mean plane and the significance of its deviation from the theoretically expected mean plane are not straightforward to determine directly from the observational sample of KBOs.   
The measurement error depends upon the number of observations available, the quality of those observations, and the distribution of those observations on the sky.  
It also depends upon the intrinsic dispersion of the orbital planes about the mean plane; for a given number of observed objects, a population with a relatively small inclination dispersion (such as the classical KBOs) will have a smaller uncertainty in the measured mean plane than a population with a wider inclination dispersion.
However there is no straightforward method to estimate the intrinsic inclination dispersion about the measured mean plane directly from our data set because the observational selection effects for the majority of KBOs in the Minor Planet Center database are poorly known.  Limiting ourselves to just the subset of KBOs from well-characterized surveys would dramatically reduce the number of available observations, especially at large semi-major axes. We therefore adopt a model distribution function for the intrinsic inclination dispersion, as described below.

Following \citep{Brown:2004b}, we estimate the measurement uncertainty of the mean plane due to the number of observed objects, their observed sky positions, and a model of the intrinsic inclination dispersion by means of Monte-Carlo simulations.
We first generate a population of KBO orbits distributed about the measured mean plane with a prescribed inclination distribution. 
We then generate a synthetic observational sample from this population, measure the mean plane of this sample, and repeat for a large number of synthetic observational samples. The distribution of the measured mean planes of these synthetic samples yields the uncertainty associated with the computed mean plane of the true observational sample.

Similarly, to assess the significance of deviations from the theoretically expected mean plane, we first generate a population of KBOs distributed about the expected mean plane with a prescribed inclination distribution. 
We generate many synthetic samples from this population (again matching the sky location distribution of the true observational sample) and measure their mean planes; we then quantify how often we would expect to find a deviation as large as that of the true observational sample if the expected mean plane is the true mean plane.

The most important assumption in these simulations is the prescribed distribution of the simulated objects' orbital planes about their mean plane. 
The distribution of KBO orbital planes has often been modeled as a Gaussian multiplied by the sine of the inclination~\citep[e.g.,][]{Brown:2001}, 
\begin{equation}
f(i) = C \sin(i)\exp(-i^2/2\sigma^2),
\label{e:fi}\end{equation}
where $C$ is a normalization constant,
$i$ is the inclination relative to a chosen reference plane (usually the ecliptic or invariable plane), and the longitudes of ascending node, $\Omega$, relative to that plane are  assumed to be intrinsically randomly distributed.
In previous studies \citep[e.g.,][]{Brown:2001,Petit:2011,Gladman:2012,Gulbis:2010}, this functional form for the inclination distribution has been used to model the intrinsic inclination distribution of the Kuiper belt, and different values of the Gaussian standard deviation, $\sigma$, have been derived for different dynamical classes of KBOs. 
We note, however, that the intrinsic distribution of $\Omega$ will not actually be uniform random if the population's mean plane deviates from the chosen reference plane plane. In these previous studies, only the distribution of the inclination, $i$, is discussed when modeling the distribution of the KBOs' orbital planes with Eq.~\ref{e:fi}; the distribution of $\Omega$ has generally not been discussed in detail, a potentially important oversight for modeling the distribution of orbital planes.

In the present work, we prescribe the distribution of the simulated objects' orbital planes as follows. We begin with the usual definition of the inclination vector,
\begin{equation}
(q,p)=(\sin i\cos\Omega,\sin i\sin\Omega),
\end{equation}
which is the ecliptic projection of the unit vector normal to the orbital plane (Eq.~\ref{e:hhat}).
For small ecliptic inclinations, $\sin i\approx i$, the inclination distribution described by $f(i)$ (Eq.~\ref{e:fi}) is the well-known Rayleigh distribution.  
A Rayleigh distribution of inclinations, together with a uniform random distribution of $\Omega$, is equivalent to the inclination vector components, $q$ and $p$, each having a Gaussian distribution of zero mean. 
In our Monte-Carlo simulations, we adopt a Gaussian distribution of $q$ and $p$, but with an important modification: that the inclination distribution is defined about a plane other than the ecliptic.  
Therefore, to generate the prescribed distribution of KBO orbital planes for our simulations, we generate the following distribution of inclination vectors:
\begin{eqnarray}\label{eq:fp}
q &=& q_0 + q_1 \nonumber  \\
p &=& p_0 + p_1,
\end{eqnarray}
where $(q_0,p_0) = (\sin i_0\cos\Omega_0,\sin i_0\sin\Omega_0)$ is the prescribed mean plane of the synthetic population, and $q_1$ and $p_1$ are random numbers drawn from a Gaussian distribution of zero mean and a prescribed standard deviation.  (In keeping with standard terminology,  $(q_0,p_0)$ can be called the forced inclination vector and $(q_1,p_1)$ can be called the free inclination vector.)
The choice of the standard deviation is based on previous studies as well as on requiring an acceptable match between the ecliptic inclinations of our synthetic KBO samples and those of the real observed KBOs.  
(These choices,described in Sections~\ref{s:ckb} and~\ref{s:distant-kb}, are different for the classical KBOs and for the more distant KBOs.)  
The error bars that we report for the measured mean plane's inclination and longitude of ascending node are derived from Monte-Carlo simulations in which $(q_0,p_0)$ is chosen to be the measured mean plane of the observed KBOs.  
To compute the significance level of the deviation of the measured mean plane from the theoretically expected mean plane (Section~\ref{s:distant-kb}), we set $(q_0,p_0$) equal to the theoretically expected plane.  
In Section~\ref{s:distant-kb}, we also consider a non-Gaussian distribution of $q_1$ and $p_1$ to assess the sensitivity of the results to the choice of prescribed inclination distribution; the non-Gaussian distribution we adopt is obtained from an empirical fit to approximately debiased observational data.

We construct synthetic data sets by selecting objects from our simulated distribution of KBOs that are in nearly the same sky locations as the real observed objects. 
Determining the sky positions of our simulated KBOs requires assignment of semi-major axes, eccentricities, longitudes of perihelion, and mean anomalies to the simulated objects, in addition to $\Omega$ and $i$ as assigned above.  
We assign to our simulated KBOs semi-major axis and eccentricity distributions similar to those of the real observed objects; this ensures that our synthetic sample will have similar biases in these elements as the real observed population. 
We then assume that the distribution of perihelion longitudes and mean anomalies are random in the range 0--$2\pi$. 
These assignments fully determine the sky position of each simulated object. 
We build up many sets of synthetic data samples by matching their sky positions to those of the real objects. By matching the synthetic samples' distribution of sky locations to that of the real observed sample we also naturally approximately account for of the inclination biases in the real observed sample.

For each prescribed distribution of orbital planes, we generate 40,000 synthetic data sets, each having the same sample size as the one real data set. 
The distribution of mean planes measured in these synthetic data sets is then the distribution from which we compute the 1--$\sigma$, 2--$\sigma$, and 3--$\sigma$ uncertainties of the measured values of $i_m$ and $\Omega_m$ as well as the significance levels for the measured deviation from the expected mean plane. 
Complete details of our simulations and our calculations of the mean plane measurement uncertainties are given in Appendix~\ref{a:sims}.

\section{Results}\label{s:results}

\subsection{The mean plane of the classical Kuiper belt}\label{s:ckb}

As of October 2016, there were 621 known apparently non-resonant main classical Kuiper belt objects, that is, all  KBOs with well-determined semi-major axes in the range 42--48 au.  
With the methods described in the previous section, we find that the mean plane of this population has J2000 ecliptic-equinox inclination and longitude of ascending node  $i_m=1.8^{\circ}$$^{+0.7^{\circ}}_{-0.4^{\circ}}$ and $\Omega_m=77^{\circ}$$^{+18^{\circ}}_{-14^{\circ}}$ (with 1--$\sigma$ uncertainty estimates).
We can compare this with the forced plane calculated with linear secular theory taking account of the known giant planets on their current orbits \citep{Murray:1999SSD}. At $a=45$~au (near the center of our sample range), that forced plane has $i_0=1.7^{\circ}$ and $\Omega_0=92^{\circ}$; this is within the $1-\sigma$ uncertainties of the measured mean plane. 

\begin{figure*}[htbp]
     \centering
    \begin{tabular}{cc}
      \includegraphics[width=2.7in]{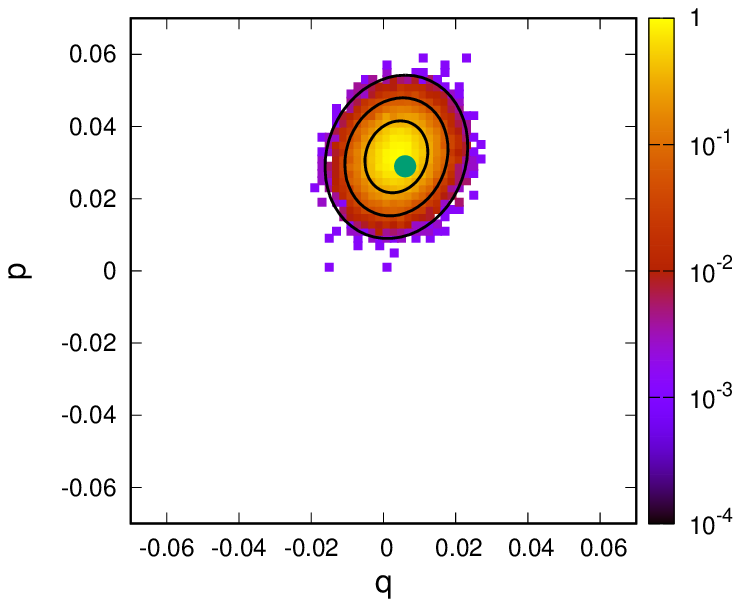} &
      \includegraphics[width=2.7in]{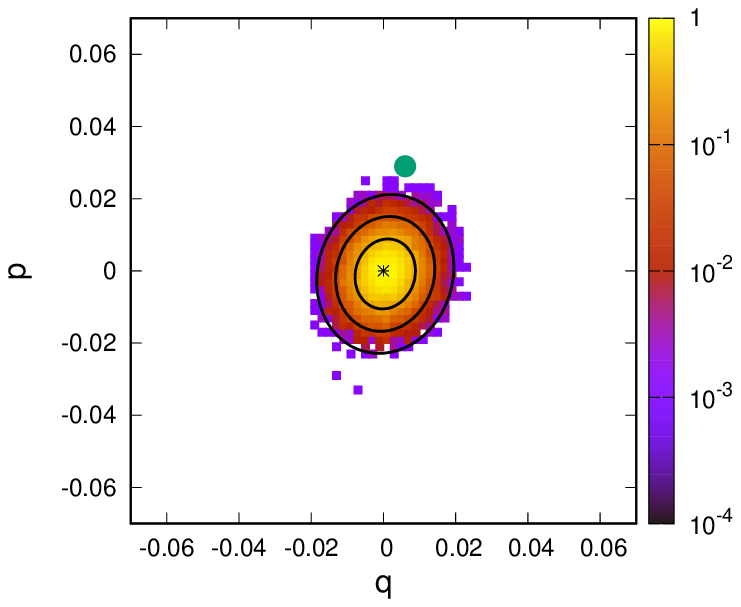} \\
      \end{tabular}
   \caption{ The mean plane of the classical KBOs (non-resonant KBOs with semi-major axis in the range 35--50 au), as measured by minimizing $\Sigma{ | \hat v_t \cdot \hat n_p |}$.  Left panel: the colored map shows the distribution of the mean plane of synthetic random samples drawn from a distribution whose forced inclination is prescribed  by linear secular theory. Right panel: the colored map shows the distribution of the mean plane of synthetic random samples drawn from a distribution whose forced inclination is defined by the ecliptic.  The mean plane of the observed sample is shown as a green dot.  The colored map is  the density of recovered planes in each pixel normalized to the maximum density point. The black ellipses represent the 1--, 2--, and 3--$\sigma$ (68.2\%, 95.4\%, and 99.7\%) limits of the expected mean plane distribution.}
   \label{f:monte-carlo-ckb}
\end{figure*}

However, linear secular theory predicts that the forced plane varies significantly with semi-major axis in the 35--50~au range that encompasses the classical belt.   To better assess whether the measured mean plane, $(i_m,\Omega_m)$, of the classical belt is consistent with linear secular theory, we carried out Monte-Carlo simulations of the population of non-resonant KBOs as follows.
We modeled the intrinsic inclination distribution of the classical Kuiper belt about its theoretically expected mean plane as a sum of two Rayleigh components,
\begin{eqnarray}
f_1(i) &\propto& \frac{\sin i}{\sigma_1^2} \exp ({-\frac{\sin^2i}{2\sigma_1^2}}), \nonumber \\
f_2(i) &\propto& \frac{\sin i}{\sigma_2^2} \exp({-\frac{\sin^2i}{2\sigma_2^2}}).
\end{eqnarray}\label{eq:ckbi}
We chose parameters $\sigma_1=\sin3^{\circ}$ and $\sigma_2 = \sin13^{\circ}$, and we forced the synthetic KBO samples to be approximately evenly split between $f_1$ and $f_2$. 
These choices provide an acceptable match (i.e. not rejectable at 95\% confidence when an Anderson-Darling test is applied) between these synthetic samples' ecliptic inclinations and the real observed ecliptic inclinations and are roughly consistent with observational constraints on $\sigma_1$ and $\sigma_2$ for the classical belt inclination distribution \citep[e.g.,][]{Brown:2001,Gulbis:2010,Petit:2011,Petit:2017}. 
We generated 40,000 simulated sets of 769 objects with the same semi-major axes as the real classical KBOs, each one was assigned a forced inclination $(q_0,p_0)$ equal to the theoretically expected forced inclination vector for their given semi-major axis,  and we assigned each one a random value of the free inclination, $(q_1,p_1)$ taken from this model inclination distribution (see Appendix~\ref{a:sims} for further details).
We then computed the mean plane of each synthetic dataset by minimizing $\sum|\hat{\bf n}\cdot\hat{\bf v}_t|$. 
The distribution of computed mean planes for the synthetic population is shown in the left panel of Figure~\ref{f:monte-carlo-ckb}.   
The colored map shows the normalized density distribution of mean planes recovered from the simulated data sets; also indicated within the density map are the 1--, 2--, and 3--$\sigma$ ellipses that encompass 68.2\%, 95.4\%, and 99.7\% of the simulated measurements, respectively. 
We conclude that the measured mean plane of the real set of observed objects (indicated with the green dot) is within 1--$\sigma$ of the secular theory prediction.

For reference (and for comparison with previous studies), we also carried out similar Monte-Carlo simulations in which the mean plane was prescribed to be the ecliptic. We find that the observational sample is inconsistent with the ecliptic as its mean plane at greater than 3--$\sigma$ significance (right panel of Figure~\ref{f:monte-carlo-ckb}). 
Simulations with the invariable plane as the prescribed mean plane of the synthetic datasets show that the observational sample is inconsistent with the invariable plane as its mean plane at greater than 2--$\sigma$ significance.

We observe in the right panel of Figure~\ref{f:monte-carlo-ckb} that the distribution of measured mean planes in our Monte-Carlo simulations is not precisely centered on the prescribed mean plane in the simulations ($p=q=0$ in this case), and the distribution also deviates slightly from circular symmetry in $(q,p)$. 
We show in Appendix~\ref{a:biases} that these deviations are owed almost entirely to the non-uniform distribution in ecliptic longitude of the observed KBOs. 
For the classical Kuiper belt, this effect is relatively small because most of the population has been detected near the ecliptic and the longitude coverage along the ecliptic has only small gaps.  
However, as shown in Section~\ref{s:distant-kb}, this bias is more pronounced in the more distant observational sample of KBOs.

\begin{figure*}[htbp]
   \centering
   \includegraphics[width=4.2in]{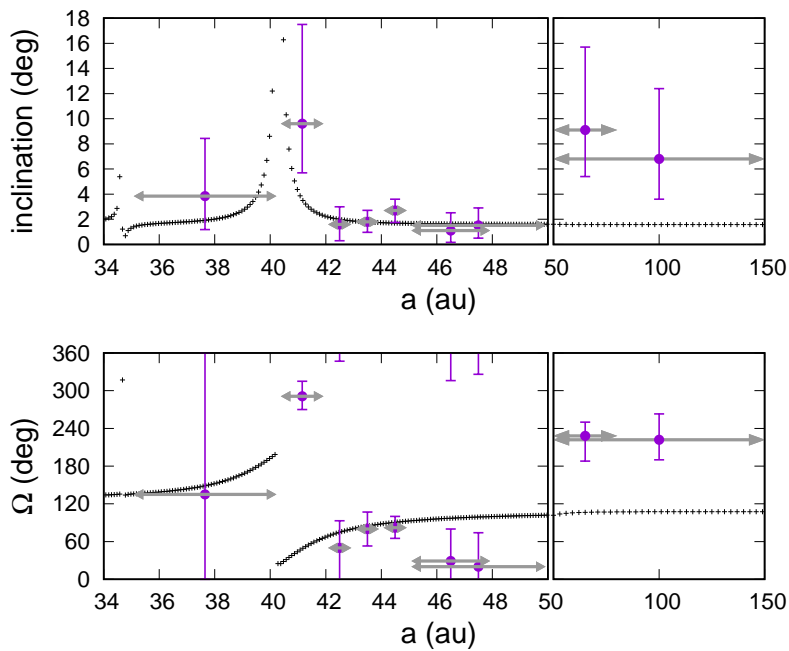}
\caption{  The plane of the Kuiper belt predicted by linear secular theory ($i_0$ and $\Omega_0$, shown in black) compared to the plane of symmetry of the velocity vectors of the observed non-resonant objects (shown in purple); the left panels show the classical Kuiper belt region while the zoomed out right panels show the more distant Kuiper belt (discussed in Section~\ref{s:distant-kb}). The horizontal gray arrows indicate the semi-major axis bin for each mean plane measurement. The vertical error bars are the 1-$\sigma$ uncertainties, obtained from Monte-Carlo simulations. }
\label{f:ckbi}
\end{figure*}

\begin{figure*}[htbp]
   \centering
   \includegraphics[width=5.2in]{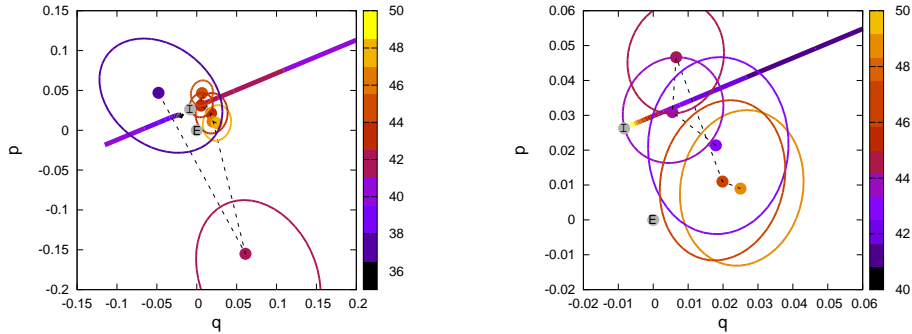}
\caption{ The plane of the Kuiper belt predicted by linear secular theory, ($q_0,p_0$), is indicated with the nearly straight but discontinuous colored line). The measured mean plane is shown as colored dots, and the 1-$\sigma$ uncertainty is indicated with colored ellipses. Both the secular theory line and the measured planes are color-coded according to semi-major axis (color bar). (The color-coded semi-major axis of the measured mean plane is set to the value in the center of the corresponding semi-major axis bin in Figure~\ref{f:ckbi}.) The left panel shows the bins from 35--50~au, while the right panel is zoomed in to show the classical Kuiper belt from 42--50~au. For reference, the ecliptic and invariable planes are indicated by gray dots labeled ``E" and``I", respectively.}
\label{f:ckbpq}
\end{figure*}

So far we have treated the classical Kuiper belt as one population. 
While the simulations above show that the entire sample is consistent with linear secular theory, the sample is dominated by KBOs in the semi-major axis range 42--45~au where the predicted plane changes very little as a function of semimajor axis.
This means that the measured mean plane of our $a<50$~au sample can also be acceptably modeled as having a mean plane equal to the linear secular theory prediction for $a=45$~au.
However, the observational sample of KBOs interior to 50~au is sufficiently large that we can subdivide it into semi-major axis bins to better test whether the observed mean plane as a function of $a$ confirms the expectation from linear secular theory, including the expected warp near 40--42~au. 
We divided the sample of non-resonant classical KBOs into 7 semi-major axis bins: 35--40.3~au (43 objects), 40.3--42~au (82 objects), 42--43~au (100 objects), 43--44~au (186 objects), 44--45~au (141 objects), 45--48~au (194 objects), and 45--50~au (217 objects).
The boundary between the first two bins is chosen to be the center of the $\nu_{18}$ secular resonance; the outermost two bins overlap because there are too few objects in the range 48--50~au for a separate bin.
For each of these semi-major axis bins, we calculated the mean plane, $(i_m,\Omega_m)$, along with the 1-$\sigma$ uncertainties from Monte-Carlo simulations with synthetic data sets.  The results are shown in Figure~\ref{f:ckbi}, together with the linear secular theory for the expected mean plane as a function of semi-major axis.  
Figure~\ref{f:ckbpq} shows the same results in $q,p$ space.

 Examining Figures~\ref{f:ckbi} and ~\ref{f:ckbpq}, the measured mean plane as a function of $a$ appears to follow linear secular theory fairly well in the classical belt region; though some deviation is apparent for the outermost bins and the bin just outside the $\nu_{18}$ secular resonance. 
The warp near $a\sim40$~au at the location of the $\nu_{18}$ nodal secular resonance is clearly evident: we see that the measured mean plane's node undergoes a dramatic transition across our innermost two bins and that the 40.3--42.0~au bin's mean plane is significantly inclined to the ecliptic.
When we compare the measured mean planes for the innermost two bins, they differ from each other at $>$3-$\sigma$ confidence. 
However, the longitude of ascending node for the bin just outside the $\nu_{18}$ does not agree with the value predicted by the linear secular theory, and the measured mean plane is inclined by $\sim13^{\circ}$ to the predicted plane for that bin; this is a nearly 3-$\sigma$ discrepancy (see the left panel in Figure~\ref{f:ckbpq}).  The cause of this discrepancy is not obvious, but we note that there is an eccentricity secular resonance, $\nu_8$, in the same vicinity, and linear secular theory does not account for the coupling between eccentricity and inclination. Such coupling could lead to additional secular resonances in this region, such as those indicated in Figure~5 of \citet{Knezevic:1991}. It is possible that higher-order secular theory is necessary to accurately model the forced inclinations of Kuiper belt objects in the vicinity of the $\nu_{18}$.  Alternatively, it is possible that this discrepancy indicates a small shift of the exact location of the $\nu_{18}$ due to unseen masses.

The mean planes measured for the three semi-major axis bins in the main classical belt (42--45~au) agree very well with the secular prediction, but the outer region of the classical Kuiper belt shows some discrepancies between their measured mean planes and the linear secular theory. The measured mean plane of the 45--48~au bin deviates from the secular theory prediction by slightly more than 1-$\sigma$, while the slightly extended 45--50~au bin deviates by $\sim$2-$\sigma$ (this is most visible in the right panel of Figure~\ref{f:ckbpq}). These deviations are less statistically significant than those presented for the more distant, $a>50$~au, population in the next section, but perhaps are indicative of the perturbation away from the expected mean plane.

\subsection{The mean plane of the more distant Kuiper belt}\label{s:distant-kb}

The observational sample of KBOs at larger semi major axes, $a\gtrsim50$~au, is too sparse in semi-major axis values to allow for the small bins in $a$ that we examined for the classical KBOs.  
To measure their mean plane, we choose two overlapping semi-major axis ranges: $50~\le~a/au~\le~80$ and $50~\le~a/au~\le~150$. 
These ranges are chosen based on the semi-major axis distribution of the observed KBOs, which is heavily weighted toward lower semi-major axes due to observational biases.  
(We do not consider KBOs with $a>150$~au because they are few in number and populate the large semi-major axis range too sparsely.)
There are 125 KBOs with $50 \le a/au \le 80$ that meet our orbit-fitting accuracy requirement and are apparently non-resonant; extending this range to $50 \le a/au \le 150$ increases this number to 162.
These objects are all included in Table~\ref{t:objects} in Appendix~\ref{a:data}.  
These two bins each have a large enough number of objects to measure their mean plane and to determine the statistical significance of its deviation from the theoretically expected plane.

For the semi-major axis bin of 50--150~au, the measured mean plane has J2000 ecliptic-equinox inclination and longitude of ascending node $i_m~=~6.8^{\circ}$~$^{+5.6^{\circ}}_{-3.2^{\circ}}$ and $\Omega_m~=~222^{\circ}$~$^{+41^{\circ}}_{-32^{\circ}}$;  the semi-major axis bin of 50--80~au has a measured mean plane with $i_m~=~9.1^{\circ}$~$^{+6.6^{\circ}}_{-3.8^{\circ}}$ and $\Omega_m~=~227^{\circ}$~$^{+18^{\circ}}_{-44^{\circ}}$. 
These can be compared to the expected plane for KBOs at $a=75$ au from linear secular theory, which has $i_0 = 1.6^{\circ}$ and $\Omega_0 =107^{\circ}$. 
(This predicted plane is very close to the invariable plane, and it changes very little over the 50--150~au semi-major axis range.)
The expected and measured mean planes are shown in the right panels of Figure~\ref{f:ckbi}; also shown are the 1--$\sigma$ uncertainties quoted above, which are computed from the Monte-Carlo simulations. 
As noted in Section~\ref{s:methods} and Appendix~\ref{a:biases}, using the sky-plane velocity vectors to measure the mean plane does not entirely remove the effects of uneven sky coverage in KBO surveys. 
The Monte-Carlo distributions of the mean planes for these two semi-major axis bins are not symmetric about the prescribed mean planes; this is why the uncertainty estimates shown in Figure~\ref{f:ckbi} are not symmetric about the measured values.  
For the Monte-Carlo simulations, we assumed that the intrinsic inclinations (relative to each prescribed mean plane) are drawn from a Rayleigh distribution of $\sin i$ with parameter $\sigma=\sin 18^{\circ}$. This choice is consistent with observational constraints of this population's inclination dispersion \citep[e.g.][]{Petit:2011,Gulbis:2010}, and it yields synthetic datasets with ecliptic inclination and longitude of ascending node distributions that are not rejected at a 95\% confidence level when an Anderson-Darling test is used to compare them to the observed population's inclination and node distributions.
We discuss other possible choices for the inclination distribution later in this section.  

\begin{figure*}[htbp]
  \centering
    \begin{tabular}{cc}
      \includegraphics[width=2.7in]{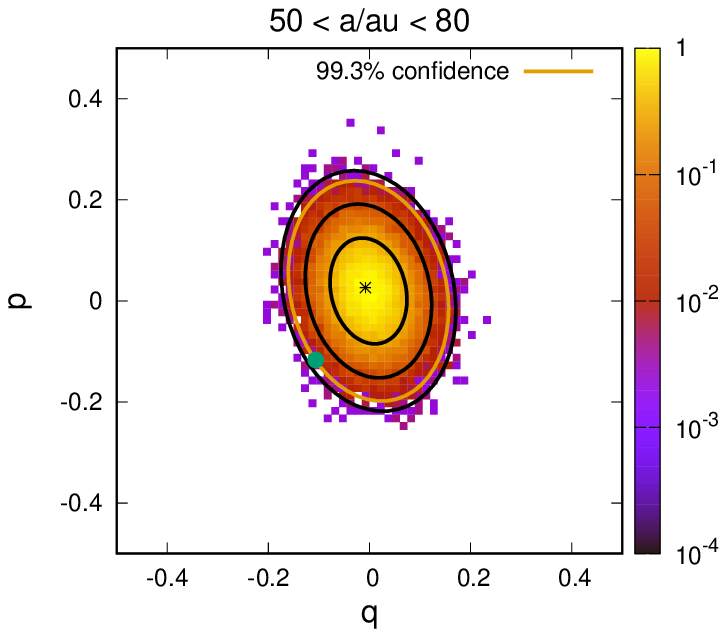} &
      \includegraphics[width=2.7in]{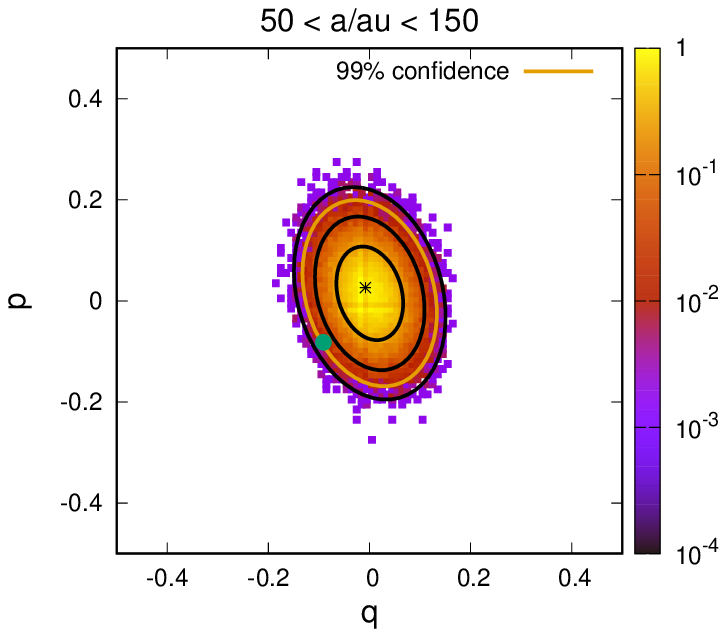} \\
      \end{tabular}
   \caption{The expected distribution of recovered mean planes (colored maps) for the distant Kuiper belt: (left panel) $50<a/au<80$ population, and (right panel) the $50<a/au<150$ population; the simulations assume a true mean plane of $i_0=1.6^{\circ}$ and $\Omega_0=107^{\circ}$ (black asterisks, the secular theory prediction for 75 au), and an inclination distribution about the mean plane described by a Rayleigh distribution with parameter $\sigma=\sin 18^{\circ}$.  The colored map is the density of recovered planes in each $p,q$ bin normalized to the maximum density. The black ellipses enclose 68.2\%, 95.4\%, and 99.7\% of the simulated mean plane measurements, representing the 1--, 2--, and 3--$\sigma$ confidence limits of the distribution.  The observed mean plane for each population is indicated by the green dots.  The confidence ellipse for each observed mean plane is shown in orange.}
\label{f:sd-plane}
\end{figure*}

The mean plane that we measure for the two samples (semi major axis ranges 50--80~au and 50--150~au) are surprisingly inclined relative to the expected mean plane. 
To assess the significance of this deviation, we repeat the Monte-Carlo simulations using the secular prediction ($i_0 = 1.6^{\circ}$, $\Omega_0 =107^{\circ}$) as the simulated population's prescribed mean plane. 
We generate simulated data sets for the two semi-major axis bins. 
The resulting distributions for the measured mean planes of the two synthetic datasets are shown in Figure~\ref{f:sd-plane}. 
We find that for both semi-major axis bins,  the measured mean plane of the observational sample is different from the theoretical prediction at the $\sim$99\% confidence level. 

\begin{figure*}[htbp]
   \centering
   \begin{tabular}{cc}
      \includegraphics[width=2.9in]{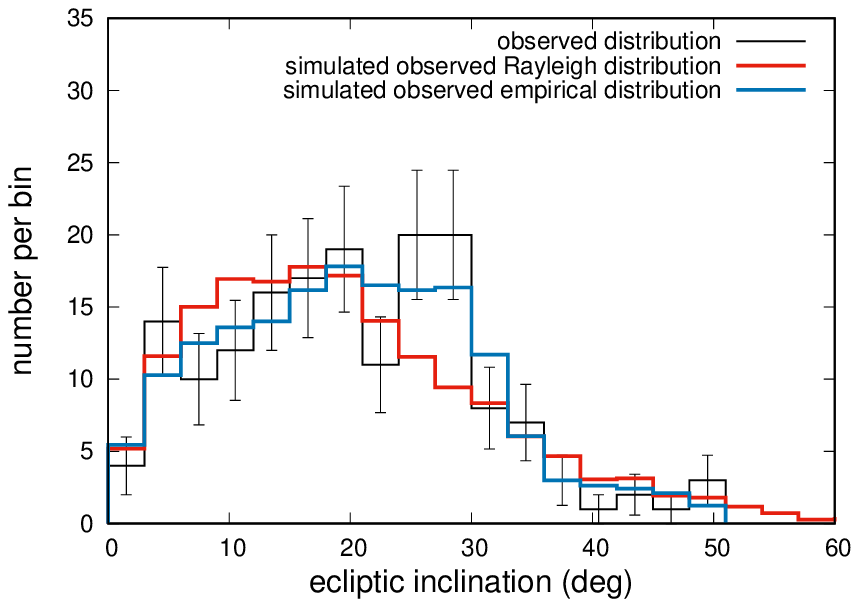} &
      \includegraphics[width=2.9in]{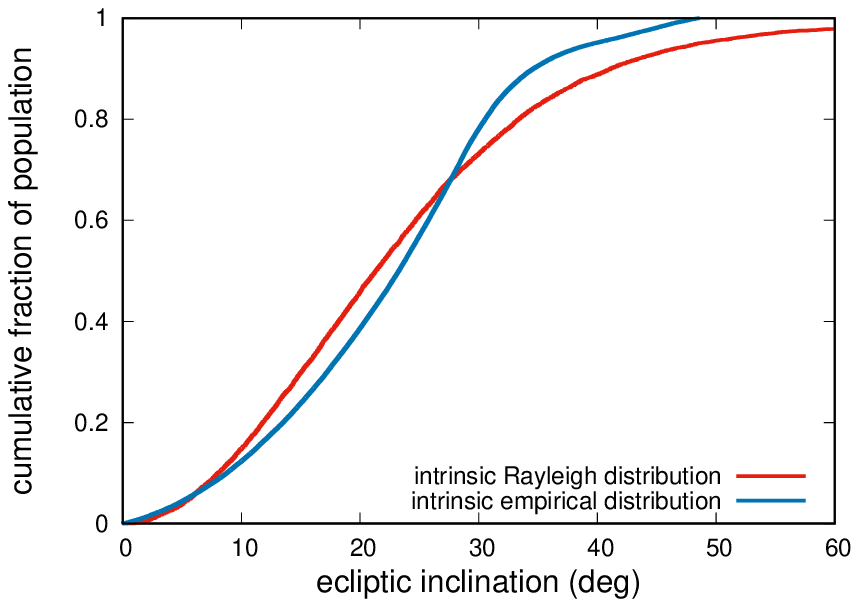} \\
      \end{tabular}
   \caption{Left panel: the observed ecliptic inclination distribution for the $50<a/au<150$ population (black histogram, poisson error bars) compared to simulated observed ecliptic inclination distributions for a population with a Rayleigh distribution in the inclination relative to the invariable plane with $\sigma=\sin18^{\circ}$ (red histogram) and from an empirically fit inclination distribution about the invariable plane (blue histogram). Right panel: cumulative intrinsic inclination distribution relative to the mean plane for the Rayleigh distribution ($\sigma=\sin18^{\circ}$, red line) and for the empirical distribution (blue line).}
   \label{f:idist}
\end{figure*}

It is important to note that the calculation of the significance level of the mean plane's deviation from the theoretically predicted mean plane is based on what we prescribed for the intrinsic distribution of inclinations about that plane. 
As noted in Section~\ref{s:uncertainties}, our measurement of the mean plane of the classical Kuiper belt has relatively small uncertainties because that population's observed inclination dispersion is fairly small. 
In contrast, the more distant observed KBOs with $a>50$~au have a much larger inclination dispersion, and their sample size is smaller. Consequently, the uncertainty ellipses in Figure~\ref{f:sd-plane} are quite large. 
The smaller sample size of this population also means that its intrinsic ecliptic inclination distribution is less well constrained than that of the classical Kuiper belt. In previous studies, the inclination distribution function has been modeled by Equation~\ref{e:fi}, with inclinations measured from either the invariable plane or the classical Kuiper belt plane; the resulting estimates for the model parameter $\sigma$ are poorly constrained and range from $\sim10^\circ$ to $\sim25^\circ$ \citep{Gulbis:2010,Petit:2011}. 
 When approximate observational biases (based on matching the observed ecliptic latitude distribution) are applied to our nominal model in which the free inclination vector components ($q_1,p_1$) have a Gaussian distribution of standard deviation $\sigma=\sin18^{\circ}$ about the invariable plane (equivalent to a Rayleigh distribution of inclinations), the resulting distribution is not statistically rejectable at $95\%$ confidence when compared to the real observations using an Anderson-Darling test. However, it over-predicts the number of observed low inclination objects and produces a narrower inclination distribution than the observed one; this is illustrated in the left panel of Figure~\ref{f:idist}, which shows this modelled distribution (in red) compared with the observed inclination distribution. A wider model distribution (larger $\sigma$) can provide a better match to the width of the observed distribution, but it then over-predicts the number of objects in the high inclination tail of the distribution.
Our nominal mean plane uncertainties assume $\sigma=\sin 18^{\circ}$ for the intrinsic distribution of inclinations of the population's mean plane because this provides statistically non-rejectable synthetic ecliptic inclination and node distributions for both the expected, lower inclination mean plane and for the observed mean plane. 
Prescribing a different width parameter for the inclination distribution, or a different functional form altogether (as \citealt{Petit:2017} have recently suggested might be necessary for the dynamically hot populations in the Kuiper belt), would result in a different estimate of the uncertainties.

To illustrate how sensitive the significance level of the mean plane deviation is to the assumed intrinsic inclination distribution, we repeated the Monte-Carlo simulations for the 50--150~au population using an empirically fit intrinsic inclination distribution about the expected mean plane. 
This empirical distribution was constructed by applying approximate debiasing factors to the observed inclination distribution, then smoothing that distribution by averaging it with a Rayleigh distribution truncated at the inclination of the highest inclination object in the real observed population; this empirical inclination distribution function is shown in blue in the right panel of Figure~\ref{f:idist}.  
(The differences with the simple Rayleigh model, shown in red, are obvious to the eye.) 
Using the empirical distribution function to assign inclinations (relative to the theoretically predicted mean plane) results in synthetic ecliptic inclination distributions that match fairly well the effective width of the real observed population's distribution without producing a high-inclination tail not seen in the real population  (Figure~\ref{f:idist}). 
Monte-Carlo simulations with this empirical inclination distribution yield a slightly wider distribution of measured mean planes of the synthetic datasets, and from these we find the significance level of the measured plane's deviation to be $\sim97\%$. 
This demonstrates that the statistical significance of the result depends on the assumed intrinsic inclination distribution.

The Monte-Carlo simulations show that two different but statistically acceptable assumptions about the intrinsic inclination distribution of the more distant KBOs find that the large deviation of the measured mean plane from the expected mean plane is statistically significant at $\sim97\%$ and $\sim99\%$ confidence.  
We caution that the exact confidence level of the deviation is model dependent. 
Additional observations and better constraints on the intrinsic dispersion of the orbital planes of the more distant Kuiper belt will be needed to confirm this deviation with higher confidence.

\section{Summary and Discussion}\label{s:disc}

Our analysis of the observed classical Kuiper belt objects finds that,  taken as a single sample, their mean plane is consistent with that predicted by the secular perturbations of the known giant planets.
The non-resonant classical KBOs with semi-major axes in the range 42--48~au have a mean plane of $i_m=1.8^{\circ}$ and $\Omega_m=77^{\circ}$ (measured relative to the J2000 ecliptic-equinox and with 1--$\sigma$ limits of 1.2--2.2$^{\circ}$ and 63--95$^{\circ}$, respectively). 
This plane is within 1--$\sigma$ of the  secular theory prediction for semi major axis $a=45$ au, and it differs from the invariable plane at the 2--$\sigma$ level. The ecliptic plane is ruled out as the classical Kuiper belt's mean plane with greater than 3--$\sigma$ significance. 
These results are consistent with \citet{Brown:2004b}, who also found the mean plane to be closer to the secular theory prediction than the invariable plane. 
However, we do not rule out the invariable plane at greater than 3-$\sigma$ significance as \citet{Brown:2004b} did, a difference likely due to different criteria for including objects in the calculation; our sample is restricted only to objects with reasonably well-determined, non-resonant orbits with semi-major axes in the classical belt semi-major axis range, while theirs included all KBOs discovered at heliocentric distances greater than 30~au. 
Our determination of this population's plane also overlaps with the results of \citet{Elliot:2005} at the 1--$\sigma$ level, though our measured mean plane is closer to the secular theory prediction than theirs. \citet{Elliot:2005} concluded that the classical Kuiper belt plane agreed better with the invariable plane than the secular theory prediction, while we come to the opposite conclusion. 
Our classical belt sample size is nearly four times larger than that available to \cite{Elliot:2005}, which likely accounts for this difference.

When we divide the $a<50$ KBOs into smaller semi-major axis bins, we find further support for the secular theory predictions (Figure~\ref{f:ckbi}). 
We clearly see a warp of several degrees in the measured mean plane near $a\sim40$~au, previously discussed by \citet{Chiang:2008}. 
The measured mean plane interior to the warp associated with the $\nu_{18}$ secular resonance agrees with the linear secular theory, but outside the warp, the measured mean plane deviates from the secular prediction by nearly $3$-$\sigma$. 
This discrepancy may be due to higher-order secular effects, such as possible coupling between inclination and eccentricity near this resonance, or perhaps due to a small shift in the exact location of the $\nu_{18}$ due to unseen masses.
Toward the outer part of the classical belt region, we also see noticeable discrepancies between the secular theory predictions and the measured mean planes for the $45-48$~au semi-major axis bin ($\gtrsim 1$-$\sigma$ discrepancy) and the $45-50$~au semi-major axis bin ($\gtrsim2$-$\sigma$ discrepancy). These discrepancies potentially signal the beginning of the more significant deviation we find in the higher semi-major axis population.

The measured mean planes of the more distant KBOs, in the semi-major axis ranges of 50--80~au and 50--150~au, are inclined to the expected mean plane by $\sim7^{\circ}$ (with a 1--$\sigma$ lower-limit of $\sim4^{\circ}$). 
Uncertainties in the model for the intrinsic inclination distribution of this population about its mean plane lead to uncertainties in the statistical significance of this surprisingly large measured deviation. 
However, it is fair to say that the measured deviation is large; for reasonable assumptions about the intrinsic inclination distribution the deviation is significant at the $\sim97-99\%$ level.

The sample of KBOs in the semi-major axis ranges of 50--150~au are thought to be affected by gravitational scattering with Neptune.  Gravitational scattering by Neptune is not expected to result in a mean plane different than the invariable plane.  
These objects are also too close to the Sun to have their angular momentum significantly affected by Galactic tides or stellar flybys \citep[see, e.g.][]{Collins:2010}. It is therefore interesting to consider possible explanations for the measured deviation from the expected mean plane of the more distant KBOs.

 In the secular theory discussed in this paper, we treat the KBOs as massless test particles. So one possible explanation for the discrepancy could be the effects of self-gravity amongst the KBOs themselves. \citet{Madigan:2016} describe a spontaneous collective tilting of a circumstellar disk of massive bodies from a state of initially nearly co-planar but highly eccentric orbits.  This mechanism requires a massive disk of KBOs, $\gtrsim 1 M_{\earth}$, much larger than estimates of ${\cal O} (0.01M_{\Earth})$ of the current mass of the Kuiper belt \citep[see, e.g.,][]{Fraser:2014}.  It also requires a speculative initial state and it is unclear that a tilted mean plane of KBOs would persist under differential orbital precession induced by the giant planets. 
Thus, the self-gravity of KBOs seems an unlikely explanation.

Another possible explanation is that an impulse-like or transient perturbation coherently altered the orbital planes of the distant KBOs in the semi major axis range of 50--150~au.
However, such a perturbation would have to have occurred recently enough that subsequent secular precession of these new orbit planes about the invariable plane has not had sufficient time to relax the KBOs' mean plane. 
The precession timescale at $a=50$~au is $\sim5$ Myr and it is $\sim15$ Myr at $a=80$~au.  This means that any transient impulse perturbation of the mean plane of objects in this semi-major axis range will tend to be erased by differential secular precession about the invariable plane on timescales of $\sim10$ Myr.
This timescale implies a perturbation much too recent to be a result of stellar flybys \citep[e.g.,][]{Levison:2004} or rogue planets \citep[e.g.,][]{Gladman:2006}.
Thus an impulse perturbation seems unlikely.

Another possibility is the presence of an inclined unseen planet in the outer solar system, which would change the secularly forced plane. 
Recent suggestions \citep{Trujillo:2014,Batygin:2016,Brown:2016,Malhotra:2016,Holman:2016a,Holman:2016b,Sheppard:2016} of a $\sim10M_\earth$ planet in the very distant solar system (beyond several hundred au) would not explain the observed deviation of the mean plane in the 50--150~au semi-major axis range. An additional $10M_\earth$ planet at $a\approx600$~au has a negligible effect on the forced plane of the Kuiper Belt at semi-major axes interior to $\sim100$~au, even if the additional planet is highly inclined. 

Achieving a significant change of the mean plane in the 50--80~au semi-major axis range requires a much closer perturber, because, absent a secular resonance, the forced plane of a KBO is nearly unaffected by an additional planet unless the KBO's semi-major axis is sufficiently near that of the planet. 
To examine the possibility that the observed change in the Kuiper belt's mean plane is due to a smaller, nearby perturber on an inclined orbit, we develop an analytical estimate based on linear secular theory for the forced inclination of a test particle as a function of the mass, semi-major axis, and inclination of such a perturber. 
In this calculation, we assume (i) that in the absence of the unseen planet, the forced plane of distant KBOs is given by the invariable plane (that is, we neglect the small difference between the forced plane  at finite semi major axes and the invariable plane as defined by the known planets), (ii) that the unseen planet's nodal precession rate is controlled by the known planets but that the unseen planet is sufficiently low mass and distant that it does not significantly affect the inclination secular modes of the known planets, and (iii) that the distant KBOs' semi-major axes are in close proximity to the unseen planet's semi-major axis.  In other words, we calculate how the invariable plane is perturbed by an inclined low mass planet in the semi-major axis range well beyond the known planets.  We then graphically solve the inverse problem of finding the combinations of unseen planet parameters (mass, semi major and inclination) for a desired forced inclination of KBOs of semi-major axis in the range of 50--100 au.  Details of this calculation are given in Appendix~\ref{a:sec-theory}.

Figure~\ref{f:m9a9} illustrates some combinations of the mass ($m_9$), semi-major axis ($a_9$), and inclination ($i_9$) of an unseen planet that could produce a forced inclination (at $a=65$~au) of the magnitude we observe for the more distant KBOs in our sample.  
(Note that in this figure, all inclinations are referenced to the local Laplacian plane determined by the known planets; for the semi-major axis range of interest here, this reference plane is very close to the solar system's invariable plane.)
Perusing this figure, we see, for example, that a Mars mass object ($\sim0.1M_{\earth}$) on a moderately inclined orbit at a semimajor axis in the range 65--80~au is sufficient to force a substantial inclination relative to the invariable plane. We show in Appendix~\ref{a:sec-theory} that such a perturber would produce significant forced inclinations over a zone of about 10--20~au width around its orbit (see Figure~\ref{f:iforced} in Appendix~\ref{a:sec-theory}).
We note that it is possible for more than one perturber to be responsible for the observed deviation.
However, the effects of a very large number of such perturbers would tend to average out and lead to a mean plane close to the invariable plane, unless their orbital planes were coincidentally aligned; thus, it is unlikely that the observed deviation can be accounted for by more than a small number of perturbers.
The semi-major axis distribution of the KBOs observed at $a>50$~au is concentrated in the lower $a$ range; nearly 80\% of the objects in our 50--150~au sample are also in the 50--80~au sample (Section~\ref{s:distant-kb}).
Thus, it is possible that the measured deviation of the mean plane could be the result of a localized perturbation due to a lower-mass planetary object currently resident amongst the scattered and scattering KBOs. Such a lower-mass perturber would be more akin to the extant planets that have been suggested to explain the population of detached KBOs within the scattered disk \citep[e.g.,][]{Gladman:2002,Lykawka:2008}.

\begin{figure*}[htbp]
\centering
\includegraphics[scale=0.5,angle=270]{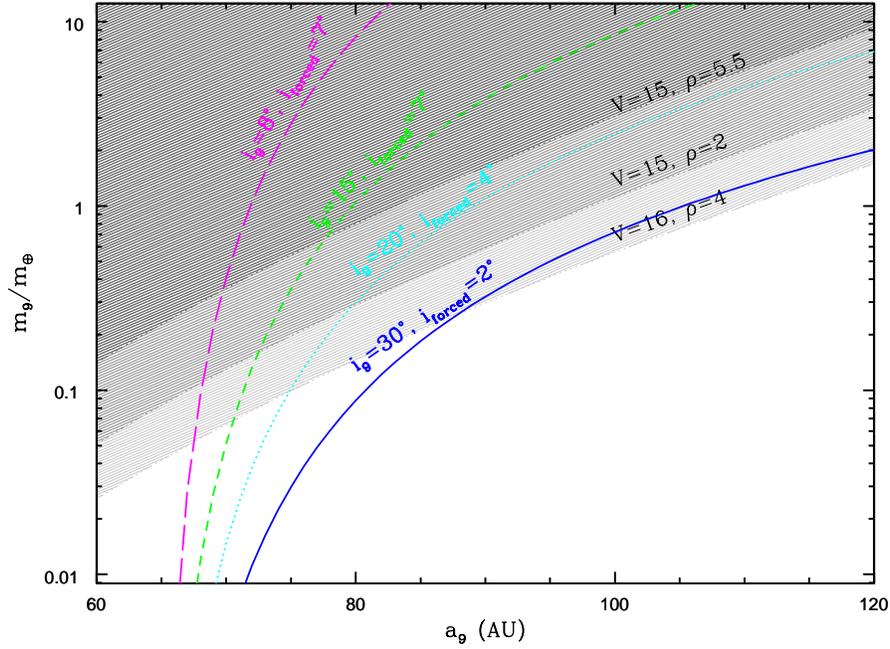}
\caption{The colored curves indicate the mass and semi-major axis combinations for a relatively close-in planet-9 with the indicated orbital inclination, $i_9$, which would produce the indicated forced inclination, $i_{\rm forced}$, of test particles at $a=65$~au (magenta corresponds to $i_9=8^{\circ}$ and $i_{forced}=7^{\circ}$, green to $i_9=15^{\circ}$ and $i_{forced}=7^{\circ}$, cyan to $i_9=20^{\circ}$ and $i_{forced}=4^{\circ}$, and blue to $i_9=30^{\circ}$ and $i_{forced}=2^{\circ}$).  The shaded zones indicate parameter regions where the planet would be brighter than the indicated visual magnitudes, assuming an albedo of 0.5 and the indicated bulk density ($\rho$ in grams per cubic centimeter). Inclinations quoted in this figure are relative to the invariable plane of the solar system.}  
\label{f:m9a9}
\end{figure*}

Figure~\ref{f:m9a9} also shows approximate visual magnitude contours as a function of the perturber's mass and distance.  For this calculation, we assume that the planet has albedo 0.5 (similar to the albedos of the icy surfaces of dwarf planets in the Kuiper belt), and density 2--5.5~g~cm$^{-3}$ (ranging from the density of dwarf planets up to the density of terrestrial planets).  We see that a perturber of Mars' mass and size in the distance range of 65--80~au would be brighter than visual magnitude $\sim17$.  Can previous observational surveys of the outer solar system rule out the existence of such an object?  It is difficult to ascertain from the literature the probability that such a perturber would have remained undetected in previous observational surveys.  We found only a brief remark in \citet{Brown:2015} stating a $\sim30$\% chance that there is one remaining KBO brighter than visual magnitude 19 within $\sim30^{\circ}$ of the ecliptic that has yet to be detected (most likely in the un-surveyed regions near the galactic plane).  There are also un-surveyed regions at higher ecliptic latitudes where relatively bright objects might remain undetected. 
It appears not impossible that a perturber on the order of Mars' size and mass, at such close distances ($\sim65$--80~au, as required to perturb the Kuiper belt's mean plane) remains to be discovered. 

It is also pertinent to note that
\citet{Holman:2016b} recently identified ranges of mass and distance combinations for an unseen planet whose perturbations could improve the orbit-fit residuals for Pluto and other KBOs.  However, their constraints do not overlap with our identified ranges of mass and distance of a perturber interior to 100~au that could produce the required inclination forcing of the distant KBOs.

The observed population of KBOs in the 50--100~au semi-major axis range does place an upper limit on the mass of such a planet residing there: the planet cannot be so massive that it would have dynamically completely cleared this region of small bodies over the age of the solar system. 
For a Mars-mass perturber, we can scale the dynamical lifetimes of Mars-crossing asteroids in the inner solar system to the longer orbital periods in the distant Kuiper belt.  \citet{Michel:2000} determined the dynamical half-lifetime of various subsets of Mars-crossing asteroids, finding half-lifetimes ranging from 45~Myr to $>100$~Myr.  Scaling these results to a Mars-mass perturber at $\sim70$~au implies a factor of $\sim300$ longer half-lifetimes, or $>13$~Gyr, for the distant KBOs; this exceeds the 4.5 Gyr age of the solar system.  Assuming that the clearing time would scale approximately inversely with the Hill radius of the planet, we can further estimate that a perturber of mass exceeding $\sim2.4\,M_\oplus$ would make the half-lifetimes of the distant KBOs to be shorter than the age of the solar system.  Thus a dynamical upper limit for the perturber's mass is $\sim2.4\,M_\oplus$.

Possibly a stronger constraint on the mass and orbit of a planet in the $\sim50-100$~au range is provided by the existence of stable populations of objects librating in Neptune's mean motion resonances. In particular, there exists a large population of objects in the 5:2 resonance at semi-major axis $a=55.4$~au \citep{Gladman:2012,Volk:2016}.  The stability of these 5:2 resonant objects has been investigated by
\citet{Lykawka:2008} who performed numerical simulations with hypothetical distant Mars-to-Earth mass planets on eccentric orbits in specific resonant configurations with Neptune. They found, for example, that a Mars-mass object on an eccentric orbit in the 3:1 resonance with Neptune ($a=62.5$~au) would destabilize the 5:2 resonant KBOs on 4 Gyr timescales, but that more distant and more massive resonant planets could allow for a surviving 5:2 resonant population if the planet's perihelion distance exceeds the aphelion distances of the 5:2 resonant KBOs.  These authors did not report on the effects of distant perturber in more general (non-resonant) orbits, and their mass/semi-major axis combinations do not fully overlap with the ranges we consider; thus it is not possible to easily extrapolate their findings for the range of parameters that we have identified for the putative perturber to account for the inclined mean plane of the distant KBOs.  We leave this to a future study.

Finally, we note the possibility that the large measured deviation of the mean plane of the distant KBOs is simply an unlikely observation, i.e., the true mean plane really is the invariable plane and that the observations thus far have unluckily sampled a set of KBOs that indicate otherwise. 
For reasonable assumptions about the distribution of orbital planes, the probability of such a statistical fluke is $1$--$3\%$.

We conclude by emphasizing that an accurate measurement of the distant Kuiper belt's mean plane provides a sensitive probe for the existence of unseen planetary-mass objects in the outer solar system.
To improve the precision of the mean plane measurement presented here requires better constraints on the intrinsic distribution of the orbit planes of the distant KBOs, which requires an increase in the sample size of well-characterized detections as well as careful modeling of the population's inclination distribution.
This modeling should treat the inclination as a vector rather than simply measuring the amplitude of the inclinations relative to the ecliptic or invariable planes. Correct modeling will help reveal deviations of the population's mean plane from the expected forced plane, will provide a dynamically meaningful representation of the distribution of KBO orbital planes, and will provide better observational constraints for models of the dynamical history of the solar system.

\acknowledgements{\noindent
We thank Scott Tremaine for discussions and comments on an early draft of this paper and Daniel Fabrycky for a helpful review. This research has made use of data provided by the International Astronomical Union's Minor Planet Center and by NASA's Astrophysics Data System.
We acknowledge funding from NASA (grant NNX14AG93G), and R.~M.~additionally acknowledges funding from NSF (grant AST-1312498).}

\appendix

\section{Kuiper Belt Data Used}\label{a:data}

We fit orbits for all of the distant objects listed in the Minor Planet Center as of October 20, 2016 using the \citet{Bernstein:2000} orbit fitting procedure.
(For objects with observational arcs longer than $\sim10$ years, we had to shorten the arc by discarding some of the older astrometry so the assumptions underlying the \citet{Bernstein:2000} orbit fitting routine were not violated \citep[see discussion in][]{Holman:2016b}. Objects with such long arcs tend to have orbits that are so well determined that the discarded astrometry has no effect on whether or not an object would be included in our sample.)
These orbits are all shown in Figure~\ref{f:a-e-i} as black crosses; we note that the excess of objects near $e=0$ in that figure is an artifact of the assumptions the \citet{Bernstein:2000} orbit fitting code makes for extremely short observational arcs.
Objects with small semi-major axis uncertainty ($da/a < 0.05$), semi-major axis $a>30$~au, and perihelion distance $q>30$~au were integrated forward under the influence of the Sun and the four giant planets for $10^7$ years using SWIFT \citep{Levison:1994} to check for orbital resonances with Neptune. 
The resulting set of nominally non-resonant Kuiper belt objects is given in Table~\ref{t:objects}, which includes the following information for each object: packed MPC designation, semi-major axis ($a$), semi-major axis uncertainty ($da$), eccentricity ($e$), inclination ($i$), longitude of ascending node ($\Omega$), argument of perihelion ($\omega$), ecliptic longitude ($\lambda$), ecliptic latitude ($\beta$), heliocentric distance ($r_h$), and the epoch of the orbit fit and sky position (in JD). 
All elements are barycentric and referenced to the J2000 ecliptic coordinate system. 
Note that the epoch for each object's orbit fit is slightly different. 
The mean plane calculation would ideally be done with all objects referenced to a common epoch, but the orbit fits use the time of the observations as the epoch. 
Where possible we have used the epoch of the most recent observations for the sky positions and orbits, so the majority of the epochs are within a $\sim5$ year span, but the full range of epochs is $\sim20$ years. 
However, typical KBOs have such slow sky motions that this difference in epochs has a minimal affect on the mean plane calculation. Most of the KBOs in our sample move only $\sim1^{\circ}$/year in longitude; their movement in latitude depends on inclination, but is typically less than a few degrees even on timescales of $\sim10$ years. 
Our estimate of the mean plane's uncertainty incorporates differences in sky position of this magnitude (see Appendix~\ref{a:sims}). 
\begin{deluxetable}{lllllllllll}[ht]
\tabletypesize{\footnotesize}
\tablecolumns{9}
\tablewidth{0pt}
\tablecaption{List objects used to calculate mean planes}
\tablehead{ \colhead{MPC Des.} & \colhead{$a$ (au)} & \colhead{$da$ (au)} & \colhead{$e$} & \colhead{$i$} & \colhead{$\Omega$} & \colhead{$\omega$} & \colhead {$\lambda$} & \colhead{$\beta$} & \colhead{$r_h$} &\colhead{epoch}} 
\startdata
15760 &  4.39316E+01 &2.012E-03 &6.93630E-02 &3.81529E-02& 6.27292E+00& 4.97942E-02&  0.526 & 0.020& 41.185 &2456571.9 \\
a3330  & 4.38022E+01& 4.304E-03& 4.76380E-02 &5.36863E-02 &2.89374E+00& 5.46686E+00 &-0.655  &0.022 &45.718 &2456450.9 \\
J94E02S &4.58115E+01 &2.933E-03& 1.15028E-01& 1.85878E-02 &2.70119E+00 &1.74919E+00 &-3.068 & 0.009 &43.617& 2457072.0 \\
K11O60B& 1.00912E+02& 7.941E-02& 6.36351E-01 &3.39501E-01& 2.49069E+00 &3.94581E+00 &-0.391 &-0.097 &38.636& 2456571.8 \\
K14J80M& 6.23061E+01& 1.901E-02 &2.62680E-01 &3.57566E-01& 3.18451E+00 &1.69196E+00 &-2.049  &0.320 &47.768 &2457162.0 \\
\enddata
\tablecomments{Table 1 is published in its entirety in the machine-readable format. A portion is shown here for guidance regarding its form and content. The MPC designations are given in their packed format (see \url{http://www.minorplanetcenter.net/iau/info/PackedDes.html}). The orbital elements are the best-fit barycentric elements from a \citet{Bernstein:2000} orbit fit to the astrometry available for each object from the MPC; $da$ is the 1-$\sigma$ semi-major axis uncertainty (taken from the orbit fit covariance matrix). All angles are given in radians, and the epoch (JD) is for both the orbit fit and sky position.}
\label{t:objects}
\end{deluxetable}

\section{Effect of observational biases on the mean plane calculation}\label{a:biases}

Computing the mean plane by simply averaging the unit vectors normal to the orbital planes does not work well for observationally biased populations. 
As discussed in Section~\ref{s:mp}, the fact that many observational surveys are performed near the ecliptic can bias the inclination of a mean plane determined in this manner.
Similarly, surveying over a limited ecliptic longitude range would yield an averaged plane with a biased longitude of ascending node, because the observational biases would favor particular values of $\Omega$ in the observed population. 
A sample of objects could yield both a biased node and a biased inclination if they were discovered in a survey with limited ecliptic latitude and longitude coverage.

\begin{figure*}[htbp]
   \centering
   \includegraphics[width=5.2in]{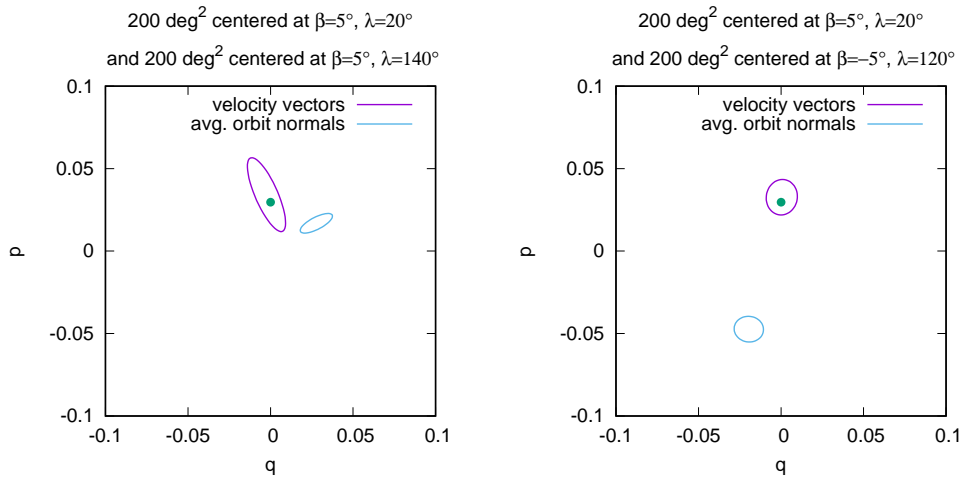}
   \caption{Mean planes determined from two hypothetical surveys of a classical Kuiper belt population with a true mean plane of $i_0=1.8^{\circ}$ and $\Omega_0=90^{\circ}$ (green dot). The purple ellipses show the 1--$\sigma$ limits on a mean plane determined by minimizing $\Sigma{ | \hat v_t \cdot \hat n |}$ for sets of simulated observations while the blue ellipses show the 1--$\sigma$ limits of the averaged orbit normals for those same simulated observations. It is clear that averaging the orbit normals of a biased observational sample is not a reliable way to recover the mean plane.}
   \label{f:biases}
\end{figure*}

Using the velocity vectors to determine the mean plane does not entirely eliminate the effects of observational biases, but it does reduce systematic errors in the calculated mean plane. 
To illustrate this, we consider two hypothetical surveys of a classical Kuiper belt population with a true mean plane of $i_0=1.8^{\circ}$ and $\Omega_0=90^{\circ}$. 
Both surveys cover 400 deg$^2$ of sky centered $5^{\circ}$ off ecliptic, with one covering two patches in longitude north of the ecliptic and the other covering one patch north and one patch south of the ecliptic at different longitudes. 
Figure~\ref{f:biases} shows what mean planes would be recovered from these two simulated surveys using either the velocity vectors (purple ellipses) or the averaged orbit normals (blue ellipses) of simulated observed objects.
It is clear that the mean planes found by minimizing $\Sigma{ | \hat v_t \cdot \hat n |}$ are much more consistent with the population's true mean plane (green dots) than those found by averaging their orbit normals.
Observational biases can change the shape of the expected distribution of the velocity vector derived mean planes (discussed in more detail below), but the velocity vector method is not subject to the same huge systematic errors evident in the averaged orbit normals. 

\begin{figure*}[htbp]
   \centering
   \includegraphics[width=3.5in]{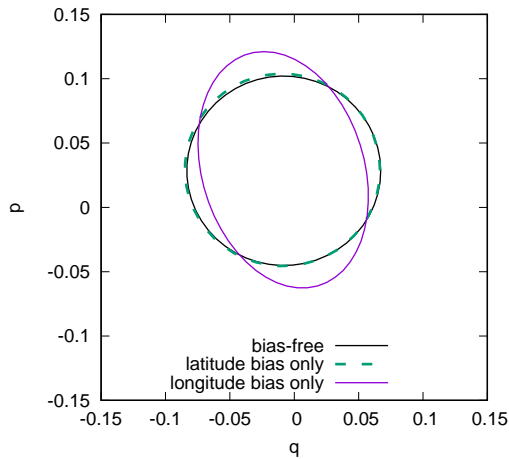}
   \caption{The 1-$\sigma$ uncertainty ellipses for the expected distribution of mean planes for a $50 \le a/au \le 150$ population of KBOs. The black circle represents the uncertainty from 162 bias-free observed objects. The green dashed circle is for 162 objects observed at the same ecliptic latitudes as the real observed objects, with no restriction on ecliptic longitude. The purple ellipse is for 162 objects observed at the same ecliptic longitude as the real observed objects, with no restriction on ecliptic latitude.}
   \label{f:ellipse}
\end{figure*}

\begin{figure*}[htbp]
   \centering
   \includegraphics[width=4in]{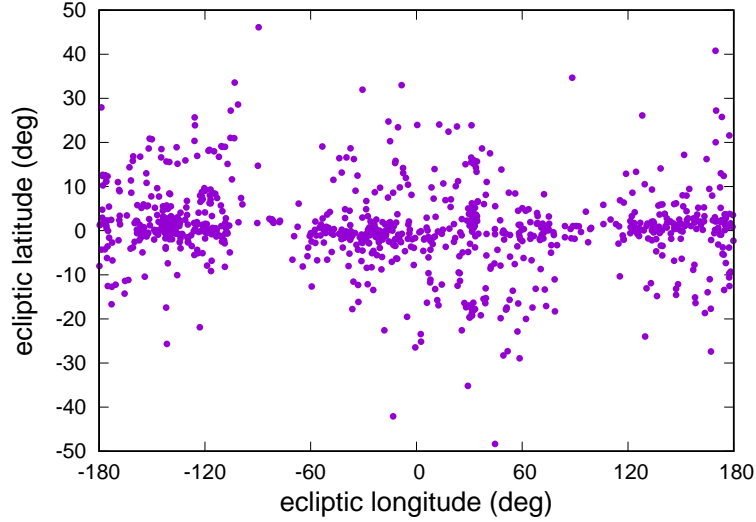}
   \caption{Sky positions for all the objects used in this work (listed in Table~\ref{t:objects}).}
   \label{f:skypos}
\end{figure*}

A set of observed objects discovered in a bias free way would have a symmetrical distribution of measured mean planes (i.e. the 1-, 2-, and 3-$\sigma$ ellipses in $p,q$ space measured relative to the true mean plane would be circles). 
Through simulations of observations, we found that biases in ecliptic latitude at discovery do not change the shape of the measured mean plane distribution, but introducing ecliptic longitude restrictions does.   
Figure~\ref{f:ellipse} shows three 1-$\sigma$ uncertainty ellipses for the expected mean plane distribution of a high-$a$ population with 162 observed objects. 
If no restrictions are placed on where objects are observed in the sky (i.e., a bias free survey), the uncertainty in the mean plane is symmetric about the true mean plane (the black circle). 
When the simulated observations are required to match the real observations in ecliptic latitude, the uncertainty is still symmetrical, but very slightly offset relative to the bias-free survey (green dashed circle). 
However, when we require that the simulated observations match the ecliptic longitude distribution of the real observations, the uncertainty ellipse becomes noticeably elongated (purple ellipse).  
The real observed population of KBOs has significant gaps in the ecliptic longitude distribution due to the galactic plane, as shown in Figure~\ref{f:skypos}. 
Almost all of the asymmetry in the mean plane uncertainties we calculate results from this ecliptic longitude bias. 
The asymmetry is less pronounced in the larger observed population of classical KBOs because most of these objects are observed near the ecliptic and the longitude coverage has only the two gaps due to the galactic plane. 
The smaller number of observed high-$a$ KBOs means their longitude coverage is less uniform; thus the asymmetry is larger. 
 
\section{Determining mean plane uncertainties using Monte-Carlo simulations}\label{a:sims}

As discussed in Section~\ref{s:mp}, we use the directions of the velocity vectors projected on the sky to determine the mean plane of a given set of KBOs.  This approach defines the mean plane as the plane of symmetry of the sky-plane projection of the velocity vectors \citep{Brown:2004b}. 
This method reduces the systematic uncertainties in the mean plane arising from observational biases; however the effects of biases cannot be entirely removed (see Appendix~\ref{a:biases}), and there are still potentially large uncertainties in the mean plane found by minimizing $\Sigma{ | \hat v_t \cdot \hat n |}$ due to the limited number of observations the sky locations of those observations, and the intrinsic dispersion of the population's planes about their mean. 
We estimate these uncertainties using Monte-Carlo simulations.

For each observed object, the following observational parameters are directly required for the mean plane calculation: inclination ($i$), longitude of ascending node ($\Omega$), ecliptic latitude ($\beta$), and ecliptic longitude ($\lambda$). 
To generate the expected distribution of observed mean planes for any given true mean plane, we simulate an intrinsic population of objects distributed in inclination about the assumed true mean plane and then select subsets of the simulated objects that are found near the sky locations ($\beta,\lambda$) of the real objects. 
For this set of simulated  ``observed" objects, we calculate the apparent mean plane by minimizing $\Sigma{ | \hat v_t \cdot \hat n |}$. 
This process is repeated 40,000 times to build a distribution of simulated ``observed'' mean planes; this is a large enough sample to define a 3-$\sigma$, 99.7\% confidence boundary for the expected mean plane.

For our simulations to determine the uncertainty in the classical Kuiper belt's mean plane, we use the following procedure to assign orbital elements to our simulated intrinsic population:
\begin{itemize}
\item semi-major axis $a$ is assigned randomly from the range $a_{obs,i}(0.99,1.01)$, where $a_{obs,i}$ is the observed semimajor axis of a randomly selected real observed object
\item mean anomaly and argument of pericenter are selected randomly from the range $(0,2\pi)$
\item eccentricity $e$ is assigned randomly from the range $e_{obs,i}(0.95,1.05)$, where $e_{obs,i}$ is the observed eccentricity of a randomly selected real observed object
\item in the simulations allowing a semi-major axis dependent mean plane, the  values of $q_0,p_0$ for the semi-major axis of the simulated object are determined from the linear secular theory of the known solar system 
\item in the simulations with a single, flat intrinsic mean plane , $q_0,p_0$ are constant for all objects
\item the free inclination vector components ($q_1,p_1$) are selected from one of the two Gaussian distributions that contribute to the classical belt inclination distribution (see Section~\ref{s:ckb} and Equation~\ref{eq:ckbi}).
\item the complete inclination vector is then calculated according to \ref{eq:fp}
\item the inclination, $i$, and longitude of ascending node, $\Omega$, are determined from the inclination vector components.
\end{itemize}
\noindent The ecliptic latitude and longitude of the object is then calculated. For each  $\beta,\lambda$ pair in the real, observed population, we repeatedly generate simulated objects until one falls in the range $\beta \pm 1^{\circ}$ and $\lambda \pm 5^{\circ}$, and then the apparent mean plane of the set of simulated observed objects is calculated; increasing or decreasing the allowed ($\beta,\lambda$) ranges within factor of about two does not significantly change the final uncertainty estimates reported in this work.  After repeating this procedure 40,000 times, we use the R statistical package to compute ellipses in $(q,p)$ that enclose 68.2\%, 95.4\%, and 99.7\% (1, 2, and 3-$\sigma$) of the simulated observed mean planes. The 1-$\sigma$ error bars in Figure~\ref{f:ckbi} correspond to the maximum and minimum vales of $i$ and $\Omega$ that fall along those ellipses for each population.

Our simulations of the high-$a$ population are similar to those for the classical Kuiper belt. The orbital elements of the underlying population are chosen in a slightly modified manner:
\begin{itemize}
\item semi-major axis and perihelion distance are assigned randomly from the range $a_{obs,i}(0.95,1.05)$ and $q_{obs,i}(0.95,1.05)$,where  $a_{obs,i}$ and $q_{obs,i}$ are the observed semi-major axis and perihelion distance of a randomly selected real observed object. 
\item mean anomaly and argument of pericenter are selected randomly from the range $(0,2\pi)$
\item the inclination vector components ($q,p$) are selected from the distribution function described in Equation~\ref{eq:fp} (assuming either a Gaussian distribution about the mean plane or an empirically fit distribution about the mean plane, see Section~\ref{s:distant-kb}).
\item the inclination, $i$, and longitude of ascending node, $\Omega$, are determined from the inclination vector components.
\end{itemize}
We again generate simulated objects until one falls near each observed $\beta,\lambda$ pair, however we also require that the object's heliocentric distance fall within 10\% of an observed object's heliocentric distance. 
We add the heliocentric distance constraint to account for the fact that the high-$a$ KBOs are more strongly biased toward discovery at perihelion than the lower eccentricity classical belt objects. 
In practice this additional constraint has only a very small affect on the error ellipses (because we are assuming a random distribution of the argument of perihelion), but we include it for completeness.

\section{Linear secular theory for the forced inclination in the Kuiper Belt with an additional distant, low-mass perturber}\label{a:sec-theory}

We consider the forced inclination vector, $(q_0,p_0)= \sin I_0(\cos\Omega_0,\sin\Omega_0)$, of a massless test particle in the distant Kuiper belt, as determined by the known eight planets (Mercury through Neptune) plus a distant planet of mass $m_9$, semi-major axis $a_9$, whose orbital plane has inclination $I_9$ to the invariable plane of the known planets.  
(Throughout this section, we use ``invariable plane" to refer to the plane which is normal to the total orbital angular momentum vector of the known planets about the Sun.)  
In the linear approximation, the forced inclination vector is straightforwardly given by the Laplace-Lagrange secular theory~\citep{Murray:1999SSD}, 
\begin{equation}
(q_0,p_0) = -\sum_{i=1}^9 \frac{\mu_i}{B-f_i}\big(\cos( f_it+\gamma_i),\sin (f_it+\gamma_i)\big),
\end{equation}
where 
\begin{eqnarray}
\mu_i = \sum_{j=1}^9 B_j I_{ji},
\qquad B = -\sum_{j=1}^9B_j,
\qquad B_j=\frac{1}{4}n\frac{m_j}{m_\odot}\alpha_j\bar\alpha_j b_{3/2}^{(1)}(\alpha_j),
\end{eqnarray}
$f_i$ and $I_{ji}$  are the frequencies and amplitudes of the nodal secular modes of the planets, 
$\gamma_i$ is a phase determined by initial conditions,
$m_\odot$ is the mass of the sun, $m_j$ and $a_j$ are the mass and semi-major axis of the $j$-th planet,
$n$ is the mean motion of the test particle,
\begin{eqnarray}
\alpha_j &=& \hbox{min}\{ \frac{a}{a_j},\frac{a_j}{a}\} < 1, \\
\bar\alpha_j &=& 
\begin{cases} 1 & \text{if $a>a_j$,} \\
                  {a_j/a} &\text{if $a_j>a$,}
\end{cases}
\end{eqnarray}
and $b_{3/2}^{(1)}(\alpha)$ is a Laplace coefficient, 
\begin{equation}
b_{3/2}^{(1)}(\alpha) = \frac{1}{\pi}\int_0^{2\pi} d\psi \frac{\cos\psi}{(1-2\alpha\cos\psi + \alpha^2)^{3/2}}.
\end{equation}
For the calculations presented below, we note the following useful approximations for $b_{3/2}^{(1)}(\alpha)$,
\begin{equation}
b_{3/2}^{(1)}(\alpha) \simeq 
\begin{cases} 3\alpha &\text{for $\alpha \ll 1$}, \\
                     \frac{2}{\pi(1-\alpha)^2} &\text{for $\alpha \longrightarrow 1$.}
\end{cases}
\end{equation}

A distant low mass planet, $a_9\gg a_8$ and $m_9\ll m_8$, would not significantly affect the linear secular solution of the known planets. Therefore, we assume that the secular modes 1--8 are unperturbed by such a planet. 
We also assume that the hypothetical ``planet-9"'s inclination relative to the invariable plane well exceeds the amplitude of the secular perturbations that would be inflicted on it by the known planets. 
In this case, the secularly forced inclination of an object in the Kuiper belt would depart from the invariable plane only for KBO's with semi-major axes $a$ close to $a_9$, and we can write
\begin{equation}
(q_0,p_0) \simeq (q_0,p_0)_0 +  S_{forced,9}\big(\cos (f_9t+\gamma_9),\sin (f_9t+\gamma_9)\big),
\end{equation}
where the first term, $(p_0,q_0)_0$, describes the forced plane defined by the known planets,
and the last term is the deviation from that plane owed to planet-9. 
For KBOs with $a>50$ au $(q_0,p_0)_0$ asymptotically merges with the invariable plane of the solar system's known planets.  
The amplitude of the perturbation of the forced plane due to planet-9 is given by
\begin{equation}
S_{forced,9} = \frac{\mu_9}{-B+f_9} 
= \frac{B_9\sin I_9}{\sum_{j=1}^8B_j +B_9+f_9}.
\label{e:siniforced}\end{equation}

The nodal precession frequency, $f_9$, of planet-9 is determined by the orbit-averaged quadrupolar gravity of all the other planets, and can be approximated as 
\begin{equation}
f_9 \simeq -\frac{n_9}{4}\sum_{j=1}^8 \frac{m_j}{m_\odot}\alpha_{j9} b_{3/2}^{(1)}(\alpha_{j9})
\simeq -\frac{3}{4} n_8\Big(\frac{a_8}{a_9}\Big)^{7/2}\sum_{j=1}^8 \frac{m_j}{m_\odot} \Big(\frac{a_j}{a_8}\Big)^2,
\label{e:f9}\end{equation}
where $\alpha_{j9}=a_j/a_9 \ll 1$, and we used the approximation $b_{3/2}^{(1)}(\alpha_{j9})\approx3\alpha_{j9}$.

For a test particle with $a \gg a_8$, we can approximate
\begin{equation}
\sum_{j=1}^8 B_j \simeq \frac{3}{4} n_8\Big(\frac{a_8}{a}\Big)^{7/2}\sum_{j=1}^8 \frac{m_j}{m_\odot} \Big(\frac{a_j}{a_8}\Big)^2.
\label{e:sumB8}\end{equation}
And, for $a$ in the semi-major axis neighborhood of planet-9\footnote{ We note that for secular theory to be valid, a test particle must be sufficiently far from the planet to avoid the effects of overlapping mean motion resonances. For a Mars-mass planet at the distances we consider, this only requires a semimajor axis difference of $\gtrsim1$~au.  }, we can approximate
\begin{equation}
B_9 \simeq \frac{1}{2\pi}n_8 \Big(\frac{a_8}{a}\Big)^{3/2}\frac{m_9}{m_\odot}\frac{\alpha_9\bar\alpha_9}{(1-\alpha_9)^2}.
\label{e:B9}\end{equation}
We note that $f_9$ has a negative value, indicating a negative rate of nodal precession (i.e., nodal regression). 
We also note that $\sum_{j=1}^8B_j+f_9$ is greater than zero when $a<a_9$ and less than zero when $a>a_9$; it vanishes when $a=a_9$, at which location the forced inclination of the test particle approaches the inclination of planet-9. 
Additionally, we note that the denominator in Equation~\ref{e:siniforced} has a singularity when $B=f_9$.   
For a low mass planet-9, this singularity (a nodal secular resonance) occurs at a semi-major axis value $a_{\nu_9} > a_9$. 
The right hand side of Equation~\ref{e:siniforced} is positive for $a<a_{\nu_9}$ and becomes negative for $a>a_{\nu_9}$; we can absorb this sign change as a phase change of $\pi$ in the forced inclination vector by replacing $\gamma_9$ with $\gamma_9+\pi$ for $a>a_{\nu_9}$, so $S_{forced,9}=\sin I_{forced}$ can be interpreted as a positive quantity for all values of $a$.

Using Equations~\ref{e:f9}--{\ref{e:B9} in Equation~\ref{e:siniforced}, we find the forced inclination,
\begin{equation}\label{eq:iforced}
\sin I_{forced} = \Big[ 1+\frac{3\pi}{2}\frac{(1-(a/a_9)^{7/2})(1-\alpha_9)^2}{\alpha_9\bar\alpha_9} \frac{\sum_{i=1}^8 m_j a_j^2}{m_9a^2}\Big]^{-1}\sin I_9.
\end{equation} 
We can re-arrange the above equation to derive an expression for the mass of planet-9,
\begin{equation}
m_9 \simeq \frac{3\pi}{2}\frac{(1-\alpha_9)^2\big(1-(a/a_9)^{7/2}\big)}{\alpha_9\bar\alpha_9} \frac{\sum_{i=1}^8 m_j a_j^2}{a^2} \frac{\sin I_{forced}}{\sin I_9 -\sin I_{forced}}.
\end{equation}

\begin{figure*}[htbp]
   \centering
   \includegraphics[width=4.2in]{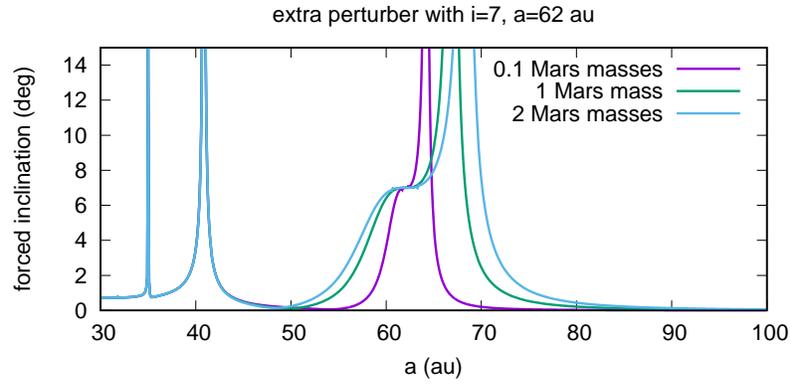}
   \caption{The forced inclinations for test particles as a function of semi-major axis from the full linear secular solution for the known giant planets plus an additional planet with $a_9=62$~au, $I_9=7^{\circ}$ and $m_9 = 0.1$~$m_{mars}$ (purple), 1~$m_{mars}$ (green), and 2~$m_{mars}$ (blue).}
   \label{f:iforced}
\end{figure*}

The forced inclinations calculated using the above approximation for the linear secular theory (Eq.~\ref{eq:iforced}) agree well with forced inclinations calculated from the full linear secular theory equations. 
Figure~\ref{f:m9a9} in Section~\ref{s:disc} illustrates a few possible combinations of planet-9 parameters, $m_9$, $a_9$ and $I_9$, that could explain the observed $\sim4^\circ$--$7^\circ$ deviation of the mean plane away from the invariable plane for KBOs with $a\sim50-80$~au. 
Figure~\ref{f:iforced} shows the forced inclinations for test particles with $a=30-100$~au given by the full linear secular solution for the known giant planets plus an additional planet with $a_9=62$~au,$I_9=7^{\circ}$ and $m_9 = 0.1,1,2$~$m_{mars}$; the range of KBO semi-major axes affected by the additional planet depends on its mass.  A Mars-mass planet can generate a $\sim15$~au wide warp in the mean plane, but even a lunar-mass object can generate a $\sim5-10$~au wide warp in the mean plane.

\end{document}